\documentclass[conference,compsoc]{IEEEtran}
\usepackage{cite}
\usepackage{url}
\usepackage{hyperref}

\ifCLASSINFOpdf
\usepackage[pdftex]{graphicx}
\else
\fi
\usepackage{parskip}
\usepackage{booktabs} 
\usepackage{multirow}

\usepackage{graphicx}
\usepackage{amsmath}
\usepackage{algorithmic}

\usepackage{array}

  \usepackage[caption=false,font=footnotesize,labelfont=sf,textfont=sf]{subfig}
\usepackage{float}
\usepackage{fixltx2e}

\usepackage{stfloats}
\usepackage{amsfonts}
\usepackage{enumitem}
\usepackage{subfig} 
\usepackage{titlesec}
\titlespacing{\section}{0pt}{*1.5}{*1.5}
\titlespacing{\subsection}{0pt}{*1.0}{*1.0}
\titlespacing{\subsubsection}{0pt}{*1.0}{*1.0}
\usepackage{parskip}
\usepackage{xspace} 

\usepackage{xspace}
\newcommand{\mypara}[1]{\noindent\textbf{#1.}\xspace}

\begin{document}
%
\title{GraphTheft: Quantifying Privacy Risks in Graph Prompt Learning}



%
\author{\IEEEauthorblockN{Jiani Zhu\IEEEauthorrefmark{1},
Xi Lin\IEEEauthorrefmark{1},
Yuxin Qi\IEEEauthorrefmark{1}, 
Qinghua Mao\IEEEauthorrefmark{1}
\IEEEauthorblockA{\IEEEauthorrefmark{1}Shanghai Jiao Tong University, China\\ \{13951230458,linxi234,qiyuxin98,mmmm2018\}@sjtu.edu.cn}
}}


\maketitle

\begin{abstract}
Graph Prompt Learning (GPL) represents an innovative approach in graph representation learning, enabling task-specific adaptations by fine-tuning prompts without altering the underlying pre-trained model. Despite its growing prominence, the privacy risks inherent in GPL remain unexplored. In this study, we provide the first evaluation of privacy leakage in GPL across three attacker capabilities: black-box attacks when GPL as a service, and scenarios where node embeddings and prompt representations are accessible to third parties. We assess GPL’s privacy vulnerabilities through Attribute Inference Attacks (AIAs) and Link Inference Attacks (LIAs), finding that under any capability, attackers can effectively infer the properties and relationships of sensitive nodes, and the success rate of inference on some data sets is as high as 98\%. Importantly, while targeted inference attacks on specific prompts (e.g., GPF-plus) maintain high success rates, our analysis suggests that the prompt-tuning in GPL does not significantly elevate privacy risks compared to traditional GNNs. To mitigate these risks, we explored defense mechanisms, identifying that Laplacian noise perturbation can substantially reduce inference success, though balancing privacy protection with model performance remains challenging. This work highlights critical privacy risks in GPL, offering new insights and foundational directions for future privacy-preserving strategies in graph learning.
\end{abstract}
\IEEEpeerreviewmaketitle
\section{Introduction}
\label{section:introduction}
Graph Neural Networks (GNNs)\cite{waikhom2021graphneuralnetworksmethods} are widely applied in practical applications such as social networks\cite{10571169,9534136}, biochemical molecules\cite{10.1093/bib/bbab513, 10.1093/bioinformatics/btac100}, and recommendation systems\cite{10.1145/3404835.3463028, 10.1145/3459637.3482092} due to their ability to effectively model complex relationships. However, they face challenges like limited labeled data\cite{Zitnik_2018}, increasing model complexity, and the need to adapt to various downstream tasks\cite{doi:10.1021/acs.jcim.7b00087}. Traditional GNN training paradigms, like `end-to-end' training and `pre-training \& fine-tuning' \cite{xu2022rethinkingnetworkpruning}, often fall short in addressing these issues.
\par
To overcome these limitations, the Graph Prompt Learning (GPL) framework \cite{10.1145/3580305.3599256,10.1145/3534678.3539249,liu2023graphprompt,NEURIPS2023_a4a1ee07,Yu2023HGPROMPTBH,Yu2023MultiGPromptFM} integrates prompt tuning\cite{ye2024promptengineeringpromptengineer, lester2021powerscaleparameterefficientprompt} into GNNs. This approach adapts models to specific tasks by using prompts to selectively focus on relevant nodes, subgraphs, or graph patterns, all while keeping pre-trained GNNs frozen. Unlike conventional GNNs, which rely on retraining or fine-tuning the model for each new task, GPL uses lightweight prompts to modify the model's behavior, enhancing efficiency and adaptability. This task-driven prompt design in GPL allows for flexible task transfer and few-shot learning, making GPL a versatile solution well-suited to evolving needs in graph-based applications.

\begin{figure}
    \centering
\includegraphics[width=1\linewidth]{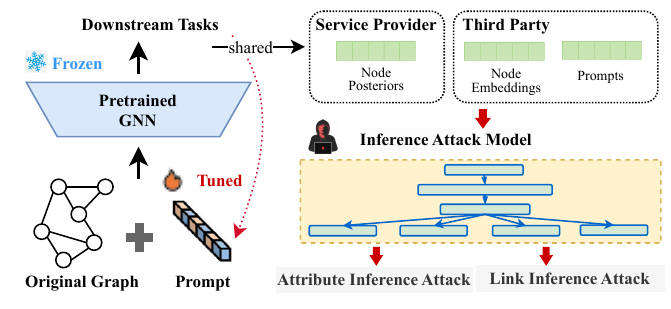}
    \caption{An overview of inference attacks on GPL. Graph prompting follows the `pre-train, prompt' paradigm by freezing the pre-trained model to adjust the prompts to the downstream task. We assume two attack scenarios: (1) When the GPL is used as a service, an attacker can launch a black box attack; (2) The attacker launches inference attacks directly against the node embeddings and prompts shared by the three parties.}
    \label{fig:Acom}
\end{figure}

\mypara{Privacy concerns}Despite these advancements, graphs are frequently considered proprietary data, often containing sensitive information such as personal user data or confidential relationships (e.g. financial transactions, etc.). While extensive efforts have been made to bolster privacy and security within traditional GNN models\cite{10.1145/3460120.3484565,10.5555/3620237.3620418,10.5555/3666122.3668088,10.1145/3589335.3651533,10.5555/3666122.3667030}, the privacy implications of GPL have remained largely unexplored. The sophisticated model architectures employed by GPL introduce additional risk dimensions due to their inherent complexity and the unique framework they utilize. In contrast to traditional GNN, GPL relies on a combination of prompts that guide the model’s attention toward specific nodes, subgraphs, or structural patterns. This reliance on prompts can create potential attack vectors, as attackers may exploit the nuances in prompt design to extract sensitive information embedded in node attributes and relationships. Thereby it is therefore necessary to examine the privacy and security loopholes in this emerging framework.

\mypara{Problem and motivation} To investigate privacy risks within the GPL framework, we envision two common application scenarios. The first is GPL as a Service, exemplified by a biomedical research institution offering genetic network analysis through a black-box API. This setup allows external users to query the model without direct access to underlying data. The second scenario is Tripartite Sharing, where, for example, a social media platform shares node embeddings and prompts generated from user data with advertisers to facilitate user classification and recommendation. While the sharing of node embeddings and prompts can enhance the accuracy of downstream graph analysis, it also introduces significant privacy concerns: (1) \textbf{GPL as a service}: When a black box query is performed on the GPL model, is the obtained node posterior information also vulnerable to inference attacks?  (2) \textbf{Tripartite sharing:} Does sharing node embeddings and prompts with third parties introduce privacy disclosure?  (3) \textbf{Prompt-driven privacy risks:} Does the use of fine-tuning prompts in the GPL exacerbate privacy risks compared to the traditional GNN model?

\begin{figure}
    \centering
\includegraphics[width=1\linewidth]{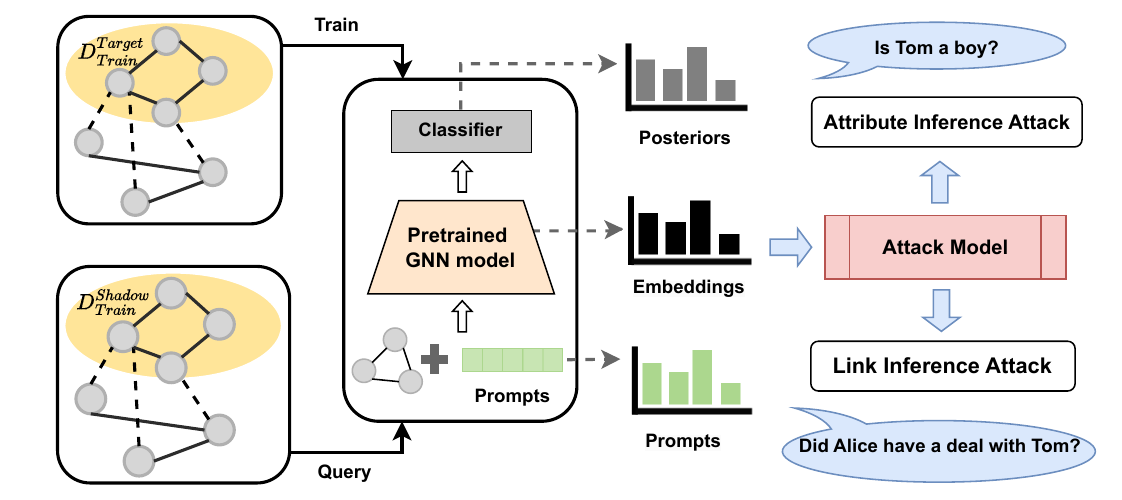}
    \caption{Schematic overview of inference attacks in GPL.}
    \label{fig:ov}
\end{figure}

\mypara{Our Contributions} In this paper, we present the first privacy risk assessment of Graph Prompt Learning (GPL), with a focus on the node classification task, as node-level privacy protection remains a critical concern in GNN-related privacy research.  We specifically target node-level inference attacks on GPL, focusing on sensitive node attributes and inter-node relationships. To achieve this, we implement Attribute Inference Attacks (AIA) and Link Inference Attacks (LIA) using established machine learning techniques and well-defined threat models. Experiments in section 4.3 and 5.3 prove that when GPL serves as the provider, attackers can use the node posterior information obtained by the black box to successfully infer some node feature information. When the GPL is faced with sharing node embeddings and prompts with third parties, it is exposed to higher privacy risks, especially node embeddings. And we can conclude by comparing the GPL with traditional GNN that tuning the tips of the GPL does not increase the risk of privacy breaches in GNN.  By identifying these risks, we aim to guide the development of more robust privacy protections in GPL applications. Our specific contributions are as follows:

\begin{itemize}
\item \textbf{First privacy risk assessment of GPL:} We perform Attribute Inference Attacks and Link Inference Attacks on GPL across three attacker capabilities, uncovering significant privacy risks and potential vulnerabilities. Notably, while the prompt itself poses privacy risks, prompt tuning in GPL does not increase privacy risks compared to traditional GNNs.

\item \textbf{Design of attribute and link inference attacks:} We introduce Attribute Inference Attack (AIA) and Link Inference Attack (LIA) to assess GPL's privacy risks. These attacks exploit training-embedded knowledge to infer private data about node attributes and graph structure. Our experiments show high success rates, with AUC scores approaching 98\% in some datasets, especially when attackers access node embeddings.

\item \textbf{Evaluation of defense mechanisms:} To counter these privacy threats, we preliminary propose a Laplacian perturbation-based defense that adds noise to node embeddings, reducing the effectiveness of both AIA and LIA. Experimental results confirm this defense effectively lowers attack success rates.

\item \textbf{Comprehensive empirical validation across datasets and prompt models:} We validate our attacks and defenses on six real-world datasets and five typical GPL methods, establishing a relatively comprehensive framework for privacy risk assessment in GPL.

\end{itemize}

\section{Preliminaries}

\subsection{Graph Neural Network} Graph Neural Networks (GNNs)\cite{waikhom2021graphneuralnetworksmethods} learn node representations by leveraging graph structures through stacked convolutional layers and message passing. Each node's features are updated by aggregating information from neighbors, allowing GNNs to encode both local attributes and the graph’s relational structure. Formally, for a graph $\mathcal{G} = (V, A, X)$, where $V$ is nodes, $A$ is the adjacency matrix, and $X$ is the feature matrix, an $L$-layer GNN iteratively computes node embeddings. At each layer $l$, a node $v$ aggregates embeddings from its neighbors:
\begin{equation} h_{N(v)}^l = \text{Aggregate}_l ({h_u^{l-1} \mid u \in N(v)}), \label{eq} \end{equation} \begin{equation} h_v^l = \text{Update}l (h_{N(v)}^l), \label{eq} \end{equation}
where $N(v)$ is the neighbors of node $v$, $h_u^{l-1}$ is the embedding of node $u$ at layer $(l-1)$, and $\text{Aggregate}_l$ is a function like mean, sum, or max. This process iteratively refines node representations for tasks such as classification and link prediction.

\subsection{Pre-trained GNN Models}
\mypara{Overview} Pre-trained models \cite{DBLP:journals/corr/abs-2106-07139} use unsupervised or semi-supervised learning techniques to capture basic patterns and structures on large-scale datasets, and these models can serve as a basis for further adaptation, fine-tuning or transfer learning for specific downstream tasks. Pre-trained models have led to significant advancements in fields such as natural language processing\cite{DBLP:journals/corr/abs-2003-08271,beltagy-etal-2019-scibert} and computer vision\cite{long2022visionandlanguagepretrainedmodelssurvey,DBLP:journals/corr/abs-2106-08254}. More recently, the success of pre-training has extended to graph data, facilitating the development of pre-trained Graph Neural Network (GNN) models\cite{xia2022surveypretraininggraphstaxonomy,hu2020strategiespretraininggraphneural}.

\mypara{DGI}  Depth Graph Infomax (DGI)\cite{veličković2018deepgraphinfomax} is an unsupervised method that captures robust node and graph-level representations by maximizing mutual information between global and local graph structures. Node embeddings are generated through GNN-based message passing to produce a graph-level summary, $\hat{g}$. DGI enhances discriminative capacity by maximizing the similarity between $\hat{g}$ and a randomly sampled negative, $\hat{g'}$.
\begin{equation} \hat{g} = \text{Readout}({h_v : v \in V}), \label{eq} \end{equation} \begin{equation} \mathcal{L}_{\text{DGI}} = -\log \sigma(\hat{g}^T \hat{g'}). \label{eq} \end{equation}

\mypara{EdgePred} Edge Prediction (EdgePred) \cite{kipf2016variationalgraphautoencoders} learns graph representations by predicting edges between node pairs. For each node pair $u$ and $v$, the model predicts an edge score $s_{uv}$, indicating the likelihood of an edge. Training with a binary cross-entropy loss function aligns the predicted edge probabilities with the actual graph structure, thereby enhancing edge prediction accuracy and robustness.
\begin{equation} s_{uv} = \sigma(h_u^T h_v), \label{eq} \end{equation} \begin{equation} \mathcal{L}_{\text{EdgePred}} = - \log ( \sum_{{u,v} \in E} s_{uv} + \sum_{{u,v} \notin E} (1 - s_{uv}) ). \label{eq} \end{equation}

\mypara{GraphMAE} Graph Mask AutoEncoder (GraphMAE) \cite{10.1145/3534678.3539321} is a self-supervised method that pre-trains GNNs by reconstructing masked node features. A subset of node features, $\tilde{\mathcal{V}} \subseteq \mathcal{V}$, is randomly masked with a [MASK] token, and the masked map is encoded to produce embeddings, denoted as $H$. During decoding, a [DMASK] token further masks parts of the encoding, yielding $\tilde{H} = \text{REMASK}(H)$. The decoder then reconstructs the original features using scalable cosine error (SCE) as the evaluation metric.
\begin{equation} \mathcal{L}_{\text{SCE}} = \frac{1}{|\tilde{\mathcal{V}}|} \sum_{v_i \in \tilde{\mathcal{V}}} \left( 1 - \frac{x_i^T z_i}{|x_i| \cdot |z_i|} \right)^\gamma, \quad \gamma \geq 1. \label{eq} \end{equation}

\mypara{GraphCL} Graph Contrastive Learning (GraphCL) \cite{10.5555/3495724.3496212} learns robust graph representations by maximizing consistency between different augmented views of the same graph. Given a graph $G$, two augmented graphs $\hat{G} \sim q(\hat{G} | G)$ are generated to form a positive pair. The GNN is pre-trained by minimizing contrastive loss, calculated based on the cosine similarity of the augmented representations. This method encourages the GNN to learn invariant features that capture the graph's essential characteristics.
\begin{equation} \mathcal{L}_{\text{sim}} = - \log \frac{\exp(\text{sim}(z_i, z_j) / \tau)}{\sum{k=1, k \neq i}^{N} \exp(\text{sim}(z_i, z_k) / \tau)}, \label{eq} \end{equation} where $z_i$ and $z_j$ represent the embeddings of the two augmented views of the same graph.

\mypara{SimGRACE} SimGRACE \cite{10.1145/3485447.3512156} is a contrastive learning-based pre-training method that deviates from traditional augmentation-based approaches. Rather than applying augmentations to the graph itself, SimGRACE introduces perturbations directly at the encoder level. This strategy enables the model to compare representations generated from distinct encoder perturbations, thereby enhancing its ability to capture semantic similarities and significantly improving its robustness against diverse data variations and noise.

\subsection{Graph Prompt Learning}
\mypara{Overview} Graph Prompt Learning (GPL) combines prompt engineering with graph learning to enable efficient adaptation of pre-trained GNN models for downstream tasks. By keeping pre-trained components frozen, GPL fine-tunes task-specific behavior using specialized prompts, allowing the model to retain generalization capabilities while adapting smoothly to new contexts.

\mypara{Differences from traditional GNN} Unlike traditional GNNs, which rely on end-to-end training or pre-training with fine-tuning, GPL leverages prompt-based adjustments that leave the pre-trained model unchanged. This approach offers a lightweight, computationally efficient way to adapt GNNs to specific tasks, particularly beneficial for task transfer and few-shot learning scenarios where traditional GNN training may struggle.

\mypara{Prompt design} The goal of graph prompt learning is to construct transformations tailored to graph structures, reframing downstream tasks to align with pre-training objectives. The ProG framework unifies diverse prompt strategies into a template. Let the frozen pre-trained graph model be $\mathcal{M}$. For a given input graph $\mathcal{G}$, a prompt module $\mathcal{P}(\mathcal{G},\mathcal{M},P,A_{\text{inner}},A_{\text{cross}})$ is learned, including a learnable prompt symbol $P \in \mathbb{R}^{K \times d}$, an internal adjacency structure $A_{\text{inner}} \in \{0,1\}^{K \times K}$, and a cross-insertion pattern $A_{\text{cross}} \in \{0,1\}^{K \times N}$. ProG\cite{zi2024prog} classifies prompts into two main types. The first, “prompt as a graph,” arranges multiple prompt symbols with an internal structure and insertion mechanism. For instance, the All-in-One\cite{10.1145/3580305.3599256} model learns insertion patterns of a set of original graphs through a learnable graph prompt module. The second type, “prompt as token,” includes methods like GPPT\cite{10.1145/3534678.3539249}, GPrompt\cite{liu2023graphprompt}, GPF, and GPF-PLUS\cite{NEURIPS2023_a4a1ee07}, where prompts are treated as individual tokens guiding the downstream task.

\mypara{Few-shot Graph Prompt Learning} Few-shot learning applies when labeled data is sparse, enabling generalization to new tasks with minimal supervision. In GPL, the prompt enables rapid adaptation of node representations, allowing the model to perform effectively with few labeled examples. During few-shot training, each class $\mathcal{C}_i$ is represented by $k$ labeled instances, where $k$ is small (e.g., $k \leq 10$). The total labeled nodes are $k \times |\mathcal{C}| = |\mathcal{Y}|$, where $|\mathcal{C}|$ denotes distinct classes. This $k$-shot framework is widely used in node-level and graph-level classification tasks.

\section{Threat Model and Attack  Capability}
\subsection{Motivation}
In this paper, we launch inference attacks on the GPL framework. Specifically, we consider a target model $\mathcal{P}$, which utilizes a pre-trained GNN, denoted as $\mathcal{M}$, along with a private dataset $D_{target}$ to generate prompts tailored to specific downstream tasks. The attacker’s objective is to infer sensitive information about the raw graph data by attacking the target model $\mathcal{P}$ within a defined capability range $\mathcal{K}$, which outlines the scope of information accessible to the attacker from the model outputs during the attack.

\subsection{Threat Model}
\mypara{Adversary's Goal} The primary objective of the adversary is to infer sensitive properties within the private dataset $D_{target}$ by exploiting information obtained during GPL.  In the context of a node-level task, the attacker seeks to deduce private attributes of individual nodes.  Focusing on a target node $v$ and utilizing the pre-trained GNN model $\mathcal{M}$, the attacker leverages prior knowledge $\mathcal{K}$ to execute privacy attacks within the GPL framework.  
\begin{equation} \mathcal{A}: v, \mathcal{M}, \mathcal{K} \rightarrow \{\text{privacy information}\}. \label{eq} \end{equation}

In these attacks, the sensitive information targeted may encompass private attributes of nodes, such as age and gender, as well as structural relationships between nodes, like connections in a social network.  Given that graph data integrates both node-specific features and structural relationships, we focus on two primary attack types: Attribute Inference Attack (AIA) and Link Inference Attack (LIA).

\begin{itemize}
\item \textbf{Attributive Inference Attack.} In GNN and GPL, node attributes serve as crucial features for constructing the graph structure. However, these attributes may contain sensitive personal information, such as age, gender, or income. The attacker seeks to infer these privatcy by analyzing outputs or intermediate representations generated by the model, potentially leading to significant privacy breaches. To this end, we use AIA to infer sensitive attributes of nodes or users from the output of models obtained for different attacker capabilities.

\item \textbf{Link Inference Attack.} In graph datasets that are rich in structural information, such as financial transactions or social networks, the interactions between nodes often signify sensitive relationships. The attacker exploits this structural data to infer private interactions between nodes. In LIAs, the objective is to deduce the existence or nature of edges between nodes, thereby potentially exposing sensitive relational information. 

\end{itemize}

\mypara{Adversary's Knowledge and Capability} We consider an inductive GPL model as the target. The attacker's knowledge and capabilities are assumed to vary based on the type of access they have to the model. First, in the most restrictive scenario, the attacker only has black-box access to the target model. This means the attacker cannot observe the internal structure of the pre-trained GNN $\mathcal{M}$ or the prompt $\mathcal{P}$. In this setting, the attacker can only query the model to obtain the posterior probabilities of nodes, which poses significant challenges and makes it the most difficult adversarial setting\cite{Salem2018MLLeaksMA}. Second, we discuss the white-box scenario, as we mentioned in the introduction of publicly available training information shared with third parties. An attacker can access more detailed information than a black box, although there are certain limitations compared to traditional white-box attacks. In this scenario, the attacker can obtain node embeddings generated from frozen pre-trained GNN $\mathcal{M}$ after GPL training, as well as trained prompt embeddings. We define three attack capabilities $\mathcal{K}$ in the two scenarios above:

\begin{itemize}
\item \textbf{Node Posterior (denoted as $P^*$):} Under the \textbf{black box} setting, the attacker can obtain the posterior probability of the node by querying the target model, which represents the direct information about the model's adaptation to the downstream task behavior.

\item \textbf{Node Embeddings (denoted as $E$):} In the \textbf{white-box} scenario, the attacker can access the node embeddings output by the pre-trained GNN $\mathcal{M}$. These embeddings reflect the model’s learned representations of nodes, providing the attacker with more substantial information about node characteristics and relationships.

\item \textbf{Prompt Embeddings (denoted as $P$):} In the \textbf{white box} scenario of GPL, the attacker can obtain the specific method of generating the prompt during the fine tuning process and obtain the prompt vector generated after the final training. These prompt embeddings may contain critical information about how to customize the model for specific downstream tasks.
\end{itemize}

We further assume that the attacker possesses a shadow dataset $D_{shadow}$, which includes its own graph structure, node features, and labels. This shadow dataset is crucial for enhancing the attacker's capabilities, as it allows them to train a surrogate model that closely mimics the behavior of the target model. Following prior work \cite{7958568}, we assume that $D_{shadow}$ is derived from the same underlying distribution as the training dataset used by the target model. This similarity ensures that the surrogate model can effectively capture the inherent patterns and relationships present in the target dataset, thereby significantly improving the attacker's ability to infer sensitive information. 

\section{Attribute Inference Attacks}
\subsection{Measurement Methodology}
When the original node features in graph data contain sensitive information, Attribute Inference Attacks (AIA) pose a significant privacy risk in GPL. To evaluate these risks, we utilize three commonly used models to construct inference attack models: Multilayer Perceptron (MLP)\cite{GARDNER19982627}, Random Forest\cite{Breiman2001RandomF}, and GraphSAGE\cite{He2021NodeLevelMI}, each offering different strengths and capabilities for inference.

\begin{figure}[htbp]
    \centering
     \subfloat[Cora]{
        \centering
        \includegraphics[width=0.4\linewidth]{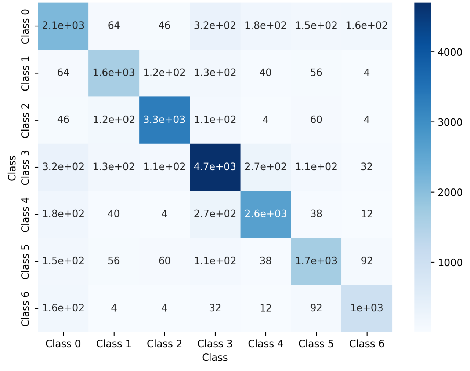}
        \label{fig:sub1}
    }
     \subfloat[Photo]{
        \centering
        \includegraphics[width=0.4\linewidth]{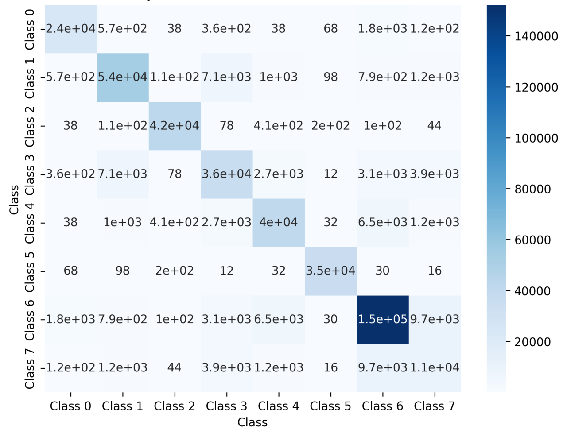}
        \label{fig:sub2}
    }
    \vspace{0.05cm}  
     \subfloat[Squirrel]{
        \centering
        \includegraphics[width=0.4\linewidth]{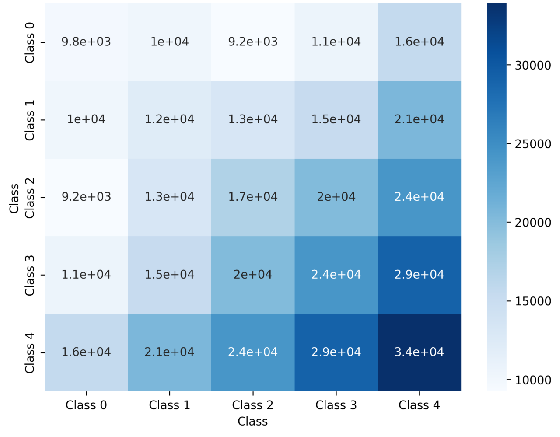}
        \label{fig:sub3}
    }
     \subfloat[Actor]{
        \centering
        \includegraphics[width=0.4\linewidth]{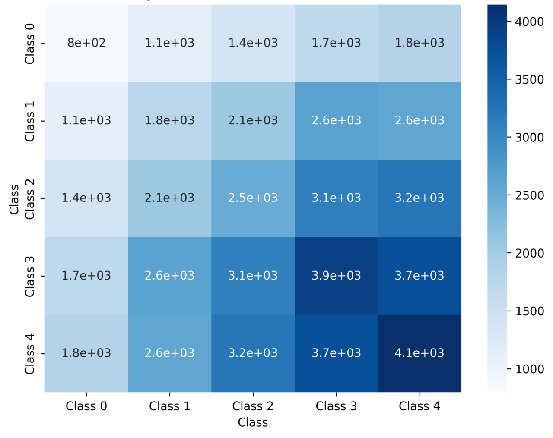}
        \label{fig:sub4}
    }
    \caption{Connectivity Matrix between Classes in Four Datasets}
    \label{fig:subfigures}
\end{figure}

\begin{itemize} 
\item \textbf{MLP-Based Attack Model:} The Multi-Layer Perceptron (MLP) is a feedforward neural network composed of multiple fully connected layers. In this architecture, each node in one layer connects to every node in the next layer. MLPs process structured input data and learn complex relationships between features.

\item \textbf{Random Forest-Based Attack Model:} The Random Forest model employs an ensemble learning approach by constructing multiple decision trees during training.  This method aggregates the predictions of these trees, enhancing predictive accuracy and stability.  The model is versatile, effectively handling various types of data, including tabular data derived from graph structures.

\item \textbf{GraphSAGE-Based Attack Model:} GraphSAGE is a specialized graph neural network architecture designed to leverage relational structures in graph data. It aggregates features from neighboring nodes, combining node attributes with local graph topology to capture contextual information effectively, making it well-suited for tasks involving graph data. Uniquely, it is the only method that requires graph structure information.

\end{itemize}

Using these models, we perform attribute inference attacks on sensitive node features. The attacker first constructs a shadow dataset $D_{shadow}$, querying the trained target model to obtain node posteriors, embeddings, or prompt vectors—depending on the attacker's level of access. These outputs are then used as input features for the attack model. The shadow dataset is split into training and test sets, with nodes and their corresponding labels from the training set used to train the attack model, thereby enabling the attacker to infer sensitive attributes.

\begin{table}[h]
\centering
\renewcommand\arraystretch{1.2}
\caption{Datasets}
\resizebox{\linewidth}{!}{
\begin{tabular}{lcccccc}
        \toprule
        \textbf{Dataset} & \textbf{Nodes} & \textbf{Edges} & \textbf{Features} & \textbf{Classes}&  \textbf{Homophily}&\textbf{Category}  \\
        \midrule
        Cora         &  2,708  & 5,429   & 1,433  & 7&0.8100&Homophilic    \\
        Photo       &  7,650 & 238,162 & 745    & 8 &0.8272&Homophilic   \\
        Computers    &  13,752    & 491,772     & 767  & 10&0.7772 &Homophilic   \\
        ogbn-arxiv   &  169,343& 1,166,243& 128   & 40&0.6551&Homophilic  \\
        Squirrel      &  5,201  & 216,933   & 2,089  & 5 &0.2239 &Heterophilic\\
        Actor        &  7,600  & 30,019  & 932    & 5  & 0.2188&Heterophilic\\
        
        \bottomrule
    \end{tabular}}
\label{dataset}
\end{table}

\begin{table*}[ht]
\small
\centering
\tiny
\renewcommand\arraystretch{0.8}
\caption{Experimental results of Attribute  Inference Attacks in 5-shot}
\resizebox{0.8\linewidth}{!}{\begin{tabular}{c|c|cccccc}
\toprule
\multirow{3}{*}{\textbf{Dataset}} & \multirow{3}{*}{\textbf{Knowledge
}} & \multicolumn{5}{c}{\textbf{Prompt Type}} \\
\cmidrule{3-8}
&  & All in one & Gprompt & GPF & GPF-plus & GPPT & w/o Prompt \\
\midrule
\multirow{3}{*}{Cora} & $P^*$ &99.58$\pm$0.14  & 88.29 $\pm$1.77 &95.01$\pm$3.69 &99.22$\pm$0.31 &16.29$\pm$5.81 &91.49$\pm$0.23\\
 & $E$ &60.24$\pm$7.12  &   99.91$\pm$0.01&97.96$\pm$0.24&96.05$\pm$0.26 &99.01$\pm$0.01&99.34$\pm$0.11\\
 & $P$ &- & - &- &86.49$\pm$2.33&-&- \\
\midrule
\multirow{3}{*}{Photo} & $P^*$ &68.79$\pm$0.14&82.61$\pm$0.41&74.30$\pm$0.20 & 81.09$\pm$0.14&85.21$\pm$0.25&82.44$\pm$0.26\\
 & $E$ &84.59$\pm$0.32 &91.00$\pm$0.14&90.73$\pm$0.16 & 91.13$\pm$0.16&90.91$\pm$0.19 &90.79$\pm$0.23\\
 & $P$ &- &-&-& 86.42$\pm$0.42&- &- \\
\midrule
\multirow{3}{*}{Computers} & $P^*$ &75.62$\pm$0.04&84.08$\pm$0.12&80.59$\pm$0.08& 77.63$\pm$0.09&50.00$\pm$0.10&83.31$\pm$0.14\\
 & $E$ &82.16$\pm$0.46 &87.16$\pm$0.19&87.00$\pm$0.31 & 86.27$\pm$0.25&87.07$\pm$0.31&87.99$\pm$0.38\\
 & $P$ &- &-&-& 80.67$\pm$0.09&-&- \\
 \midrule
 \multirow{3}{*}{ogbn-arxiv} & $P^*$ &74.58$\pm$0.21 &  81.05$\pm$0.04 &81.55$\pm$0.05&81.88$\pm$0.04&79.35$\pm$0.11 &81.03$\pm$0.04\\
  & $E$ &76.32$\pm$0.36 &83.78 $\pm$0.84 &84.73$\pm$0.57 & 82.92$\pm$0.16&83.88$\pm$0.32 &84.30$\pm$0.19\\
  & $P$ &- & - &-& 89.58$\pm$0.09&- &-\\
 \midrule
 \multirow{3}{*}{Squirrel} & $P^*$ &82.44$\pm$0.97&75.37$\pm$1.59&74.85$\pm$7.53 & 84.91$\pm$2.57&65.57$\pm$0.84&86.83$\pm$0.57\\
 & $E$ &46.72$\pm$2.66 &85.47$\pm$4.58&76.92$\pm$4.56 &75.31$\pm$3.01&83.62$\pm$6.84&65.71$\pm$2.61 \\
 & $P$ &- &-&-& 64.48$\pm$0.22&-&-\\
 \midrule
 \multirow{3}{*}{Actor} & $P^*$ &43.31$\pm$0.70 &90.98$\pm$0.31&96.98$\pm$0.54& 88.70$\pm$1.50&39.60$\pm$5.87 &90.81$\pm$0.62\\
 & $E$ &13.73$\pm$1.84 &99.93$\pm$0.01&99.72$\pm$0.15 & 97.02$\pm$1.19&99.94$\pm$0.01&99.77$\pm$0.21\\
 & $P$ &- &-&-& 99.83$\pm$0.09&-&-\\
\bottomrule
\end{tabular}}
\label{att}

\end{table*}

\subsection{Measurement Settings}
\mypara{Dataset} To evaluate the privacy leakage risks of GPL across different types and sizes of datasets, we conducted experiments on six representative datasets for node-level tasks, as summarized in Table 1. The datasets include homophilous datasets (e.g., Cora), non-homophilous datasets (e.g., Actor), and large-scale datasets (e.g., ogbn-arxiv). Figure \ref{fig:subfigures} presents an analysis of the differences between homophilous datasets (Cora, Photo) and heterophilous datasets (Actor, Squirrel) from the perspective of inter-class connection matrices. In heterophilous datasets, the inter-class correlations are significantly higher compared to homophilous datasets.

\mypara{Dataset Configuration} For each dataset, we trained a target model. Specifically, we randomly selected 20\% of the nodes and used their ground truth labels to train the target model. To accommodate few-shot learning tasks, we ensured that the number of nodes per label satisfied the training requirements during the node extraction process. To simulate and fairly evaluate different attacks, the remaining 80\%  of the nodes were compiled into a shadow dataset, $D_{shadow}$, which includes both node attributes and structural characteristics, as well as their corresponding labels. The shadow dataset was randomly split into an attack training set and an attack test set. We queried the trained target model using nodes from the shadow dataset, and the attack training set was used to train our attack model, while the test set was used to evaluate the attack's performance. Based on common sense, we assume a node-sensitive property corresponding to each dataset, consistent across all experiments.

\mypara{Metric} The primary metric for evaluating the attacks is the Area Under the Curve (AUC), a widely used measure for binary classification that is robust to class imbalance, making it more reliable than accuracy in such cases. Accuracy (ACC), which reflects the proportion of correctly classified samples, is also used as a simpler, intuitive measure, particularly valuable when data distribution is more balanced. Thus, we include ACC as an additional metric to assess the graph prompt model's performance.

\mypara{Benchmark Methods} We utilized ProG\cite{zi2024prog}, a recently proposed unified graph prompt library, as our baseline. ProG incorporates various pre-training methods at different levels, including DGI\cite{velickovic2018deep}, GraphMAE\cite{10.1145/3534678.3539321}, EdgePred\cite{kipf2016variationalgraphautoencoders}, GraphCL\cite{10.5555/3495724.3496212}, and SimGRACE\cite{10.1145/3485447.3512156}, all of which are widely used in graph learning tasks. ProG seamlessly integrates these methods into a unified graph prompt module to freeze and fine-tune pre-trained graph models, utilizing the following prompt methods:

\begin{itemize}
\item \textbf{All-in-One\cite{10.1145/3580305.3599256}:} This approach abstracts tasks into a graph-level task by extracting surrounding subgraphs from the original graph to learn insertion patterns. A learnable graph prompt module adapts these patterns, ensuring broad applicability at the graph level.

\item \textbf{GPrompt\cite{liu2023graphprompt}:} GPrompt introduces prompt vectors through element-wise multiplication within the graph, embedding task-specific prompts directly into the feature space. This approach tailors representations for specific tasks, enhancing model adaptability and performance across diverse downstream applications.

\item \textbf{GPF\cite{NEURIPS2023_a4a1ee07}:}  GPF trains a common prompt across all nodes in the original graph to enhance feature representation, while \textbf{GPF-plus} builds on this by introducing personalized prompt tokens for each node, balancing diversity and task specificity in the prompts.

\item \textbf{GPPT\cite{10.1145/3534678.3539249}:} GPPT defines a graph prompt as an additional token, including task and structure tokens, aligning with both task requirements and graph structure.

\end{itemize}

To evaluate the impact of prompt tuning in GPL on privacy risks, we define a baseline condition, \textbf{w/o Prompt}, in which prompt tuning is not applied. Instead, while keeping the pre-trained model frozen, the downstream task adapts by directly training task-specific parameters. By comparing this setup with five other prompt-based mechanisms, we assess privacy threats in the absence of a prompt mechanism.

\mypara{Implementation} For the attack target model baseline, we referenced the unified graph prompt framework and settings provided in ProG\cite{zi2024prog}. The hidden dimension of the GNN was set to $128$, and the number of layers was set to $2$. For the attack models, we set the learning rate to $1 \times 10^{-3}$ and the batch size to $64$. All attack experiment results were obtained with 95\% confidence intervals from $10$ tasks. Based on convergence behavior, we trained for $100$ epochs on all datasets and methods. All experiments were conducted on four NVIDIA GeForce RTX $3090$ machines, each equipped with $24$ GB of RAM.

\subsection{Measurement Results}
\subsubsection{Evaluations of Attack Models}
We first evaluate the performance of three attack models—MLP, Random Forest, and GraphSAGE—across different datasets to assess their effectiveness in inference attacks. As illustrated in Figure \ref{fig:com}, we conducted inference attacks using the node posterior $P^*$ on both the Cora and Actor datasets (with additional comparisons using node embeddings $E$ provided in the \textbf{Appendix}). Random Forest consistently underperformed, especially on Cora, likely due to its limitations with complex, graph-based data. In contrast, the neural models MLP and GraphSAGE showed strong and stable attack performance across prompt types. Among these models, MLP demonstrated particular robustness, as it operates independently of graph structure, unlike GraphSAGE, which depends on structural information from the graph.

Considering the more extensive attack hypothesis and attack capability, we choose MLP as the main attack model for further analysis.

\begin{figure}[H]
    \centering
\includegraphics[width=0.95\linewidth]{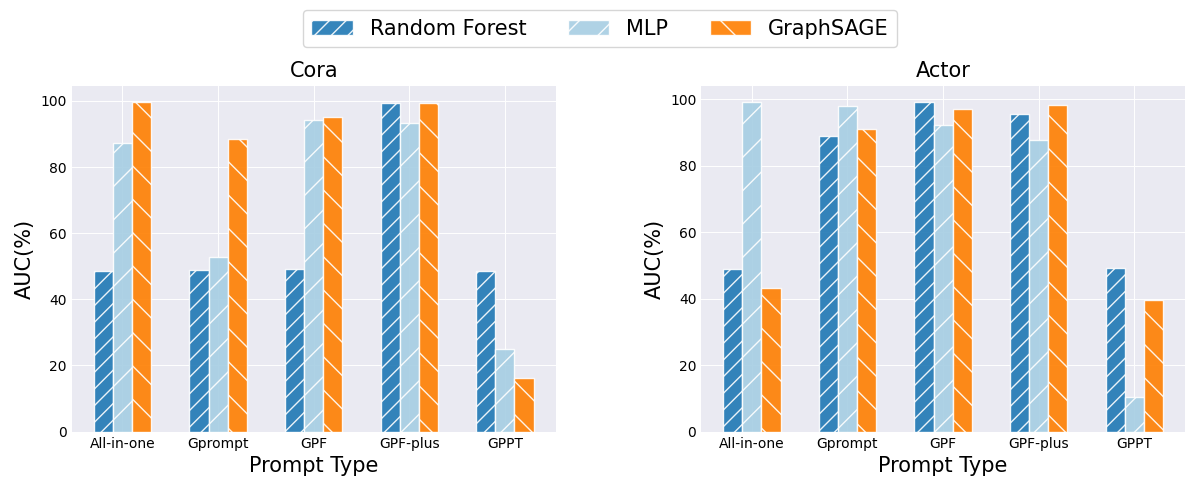}
    \caption{Attack performance of different attack models on node posteriors}
    \label{fig:com}
\end{figure}

\subsubsection{Evaluations of different attacker capabilities}
We evaluated the AIA performance of five representative graph prompt methods across three attacker capabilities. Especially, for the prompt attack, we test GPF-plus due to its unique node-wise approach among GPL models, which generates distinct, independent prompts for each node, while other methods are graph-wise. Experiments are conducted on four homophilous datasets (Cora, Photo, Computers, ogbn-arxiv) and two  heterophilous datasets (Squirrel, Actor). For consistency, we use GraphCL to pre-train the GNN model across all methods and focus on the 5-shot scenario, which is commonly used in few-shot learning (additional results for 10-shot attacks are included in the \textbf{Appendix}).

\mypara{Vulnerability of Node Posteriors to AIAs} We first assess the privacy risk associated with obtaining node posterior information via a black-box model under low attacker capability. The results indicate that, although the posterior output poses a significant privacy risk, its vulnerability is generally lower than that of node embeddings. The posterior output primarily reflects the final model prediction for specific tasks, such as classification or regression, and is more task-oriented. Being low-dimensional and containing limited structural and attribute data, these outputs are less susceptible to inference manipulation. This finding aligns with current trends in GNN attack research. 

\begin{figure}[t]
    \centering
\includegraphics[width=0.95\linewidth]{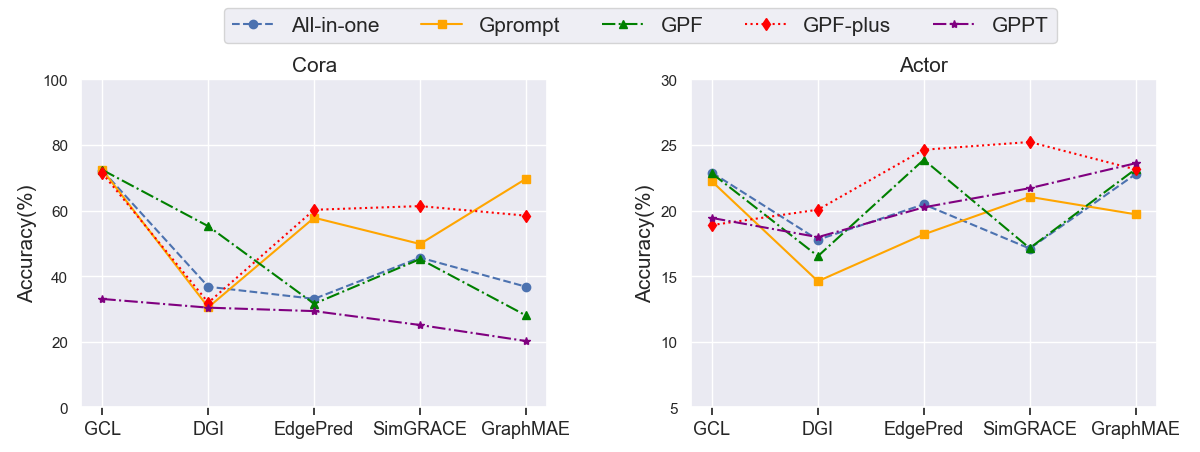}
    \caption{Performance of five GPL models under different pretraining methods}
    \label{fig:ori}
\end{figure}
\begin{figure}[t]
    \centering
    \includegraphics[width=0.95\linewidth]{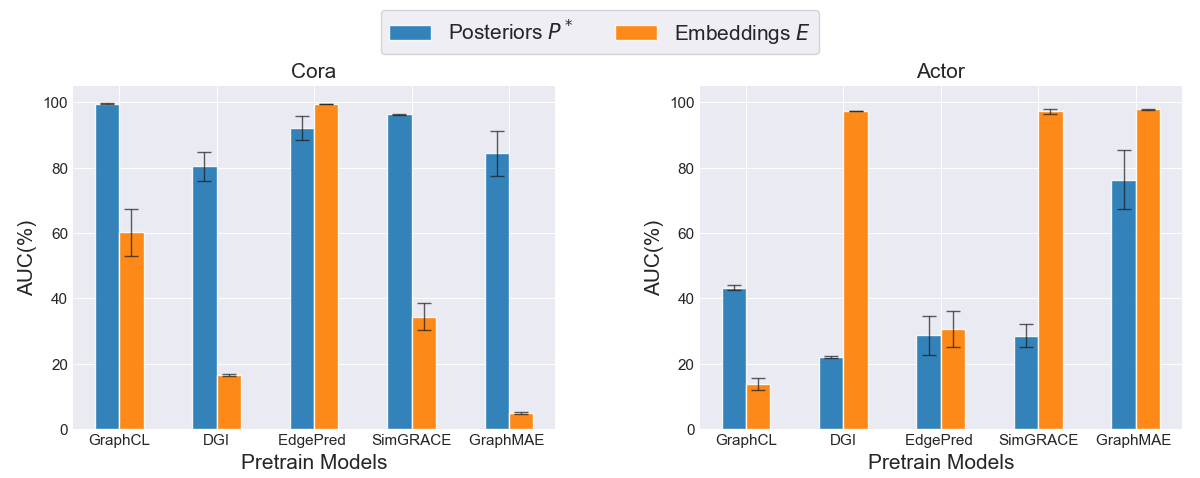}
    \caption{The performance of Attribute Inference Attack under different pretraining methods (take 'All-in-one' as an example)}
    \label{fig:a-p}
\end{figure}

However, exceptions arise, particularly with the \textbf{All-in-one} method. Our analysis of GPL methods shows that \textbf{All-in-one} employs a unique ``prompt graph" to guide the representation of the entire graph, shifting focus from individual nodes to the global graph level. This prompt-driven structure reduces emphasis on node embeddings while increasing reliance on global prompt representations. Consequently, in inference attacks targeting posterior output, such models show a higher likelihood of directly leaking original graph information. In contrast, methods like \textbf{GPPT}, \textbf{GPrompt}, and \textbf{GPF} treat prompts not as a complete graph but as tokens on individual nodes. For example, \textbf{GPrompt} uses feature inner product, while \textbf{GPF} applies generalized embedding enhancements as prompts. Here, node embeddings integrate an additional layer of the original graph’s information, increasing the risk of leakage. Due to their lower dimensionality, posterior outputs retain less sensitive information, potentially obscuring the full impact of the attack.

Therefore, based on these findings, the privacy and security of node posterior information, which is the most accessible information for attackers from the black box model, remains a significant concern and requires vigilant attention and protection.

\mypara{Privacy Risks of Node Embeddings and Prompts}  We separately explore the privacy threats when shared nodes are embedded and presented to third parties. First of all, in an attack against node embedding, particularly \textbf{Gprompt} and \textbf{GPF}, exhibit a high degree of vulnerability, with AIAs achieving a minimum of 75\% accuracy and approaching 90\% on homophilous datasets. Node embeddings inherently encapsulate the original graph structure and node attributes, embedding comprehensive context information and structural relations of adjacent nodes within a high-dimensional vector space. This direct representation heightens the probability of an attacker recovering sensitive data from the original graph and significantly impacts the feature representation in downstream layers, thereby amplifying the model's overall susceptibility to attacks.

\begin{figure}[t]
    \centering
    \includegraphics[width=0.95\linewidth]{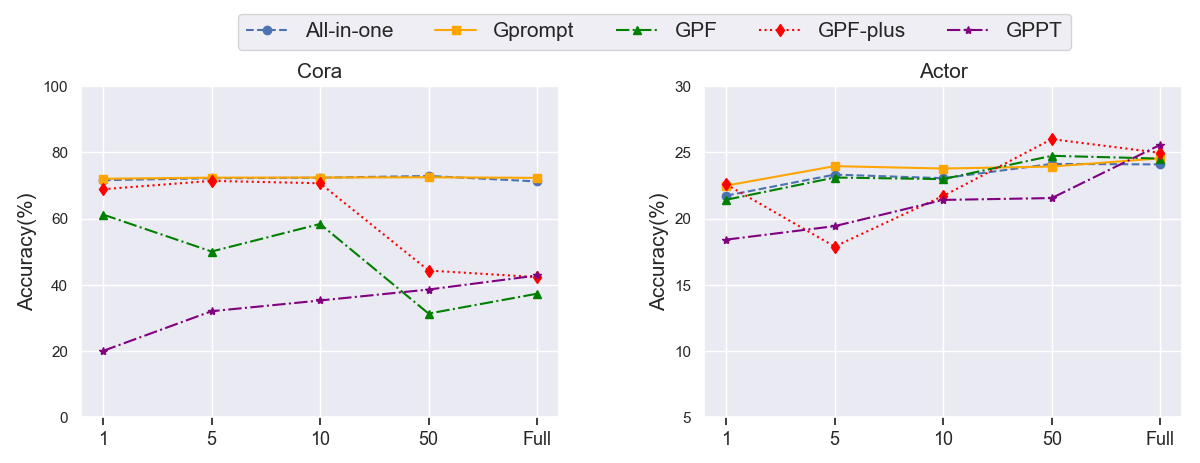}
    \caption{Performance of five GPL models with different $k$-shot Settings}
    \label{fig:ori-k}
\end{figure}
\begin{figure}[t]
    \centering
\includegraphics[width=0.95\linewidth]{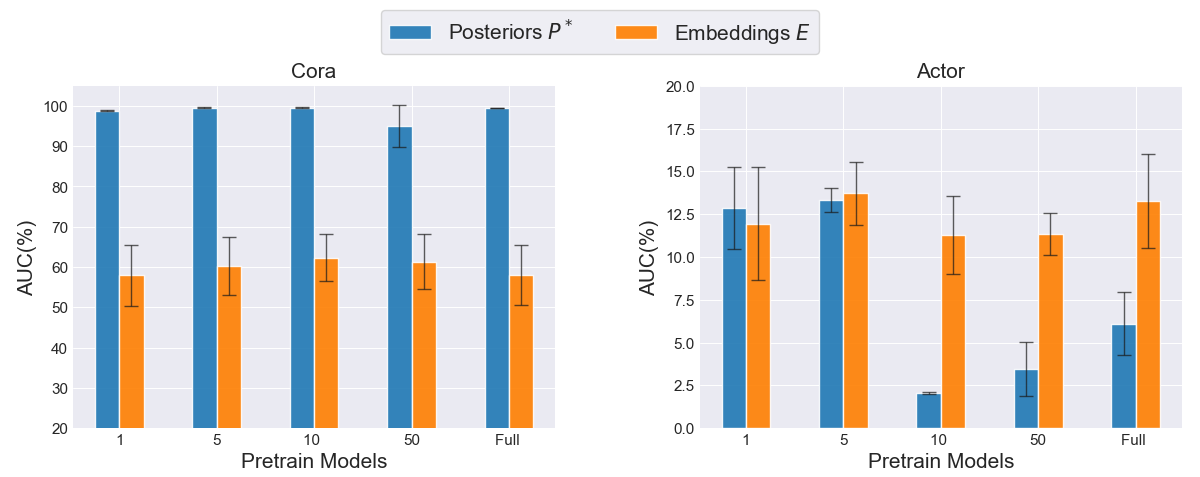}
    \caption{The performance of Attribute Inference Attack under different k-shot settings (take 'All-in-one' as an example)}
    \label{fig:att-k}
\end{figure}

Secondly, to ascertain whether shared generated prompts pose an equivalent privacy threat, we concentrate on \textbf{GPF-plus}. In \textbf{GPF-plus}, prompts for each node are specifically tailored to capture its unique characteristics, enhancing the personalization of node representation. Attack results indicate that AIA launched on the obtained prompts can effectively extract the feature representation of individual nodes, with attack efficacy reaching up to 99\% in the Actor dataset. Although this approach shows slightly less vulnerability compared to other embedding-based methods, the privacy risks are still significant and warrant careful consideration. This insight is particularly crucial for safeguarding copyright security in future prompt-based applications.

In light of these findings, we conclude that in GPL, sharing node embeddings and prompts may pose a significant risk of node privacy disclosure, as attackers can potentially recover sensitive attributes from the original graph.

\begin{table*}[ht]
\small
\centering
\tiny
\renewcommand\arraystretch{0.8}
\caption{Experimental results of Link Inference Attacks in 5-shot}
\resizebox{0.8\linewidth}{!}{\begin{tabular}{c|c|cccccc}
\toprule
\multirow{3}{*}{\textbf{Dataset}} & \multirow{3}{*}{\textbf{Knowledge
}} & \multicolumn{5}{c}{\textbf{Prompt Type}} \\
\cmidrule{3-8}
&  & All in one & Gprompt & GPF & GPF-plus & GPPT&w/o Prompt  \\
\midrule
\multirow{3}{*}{Cora} & $P^*$ &94.77$\pm$0.15&94.64$\pm$0.09&92.32$\pm$0.22 & 91.11$\pm$0.11&96.46$\pm$0.16&92.24$\pm$0.14 \\
 & $E$ &96.66$\pm$0.10 &97.90$\pm$0.30&97.77$\pm$0.07& 98.22$\pm$0.05&97.62$\pm$0.04&97.91$\pm$0.14\\
 & $P$ &- & - &- &62.69$\pm$2.19&-&- \\
\midrule
\multirow{3}{*}{Photo} & $P^*$ &83.85$\pm$0.07&88.81$\pm$0.13&87.03$\pm$0.08 & 89.58$\pm$0.08&89.46$\pm$2.14&92.16$\pm$0.09\\
 & $E$ &93.26$\pm$0.43 &97.75$\pm$0.08&97.72$\pm$0.08 & 97.55$\pm$0.16&97.66$\pm$0.16&97.89$\pm$0.36 \\
 & $P$ &- &-&-&74.33$\pm$0.25&-&-  \\
\midrule
\multirow{3}{*}{Computers} & $P^*$ &81.06$\pm$0.06&86.52$\pm$0.96&84.60$\pm$0.07 & 87.46$\pm$0.07&69.26$\pm$8.13&87.91$\pm$0.15\\
 & $E$&84.94$\pm$1.04 &95.31$\pm$0.81&94.85$\pm$0.23 & 95.86$\pm$0.22&95.52$\pm$0.52&96.59$\pm$0.38 \\
 & $P$ &- &-&-& 73.65$\pm$0.11&-&- \\
 \midrule
 \multirow{3}{*}{ogbn-arxiv} & $P^*$  &91.53$\pm$0.03&94.32$\pm$0.25&94.96$\pm$0.09 & 94.27$\pm$0.14&91.94$\pm$1.12&93.34$\pm$0.09\\
 & $E$ &93.32$\pm$0.06&97.92$\pm$0.16&97.98$\pm$0.01 &97.94$\pm$0.11&97.52$\pm$0.03& 98.86$\pm$0.01\\
 & $P$ &- &-&-& 77.82$\pm$2.15&-&- \\
 \midrule
 \multirow{3}{*}{Squirrel} & $P^*$ &78.71$\pm$0.34 &92.32$\pm$0.16&93.29$\pm$0.02 & 90.54$\pm$0.04&94.58$\pm$0.04&94.20$\pm$0.16\\
 & $E$ &89.64$\pm$0.12&97.59$\pm$0.02&97.64$\pm$0.04 & 96.15$\pm$0.46&97.65$\pm$0.03 &97.89$\pm$0.09\\
 & $P$ &- &-&-& 68.76$\pm$0.28&-&-\\
 \midrule
 \multirow{3}{*}{Actor} & $P^*$ &66.74$\pm$0.35 &80.03$\pm$0.06&75.93$\pm$0.15 & 77.21$\pm$0.15&86.92$\pm$0.24&80.02$\pm$0.18 \\
 & $E$&73.09$\pm$0.25 &92.94$\pm$0.11&93.31$\pm$0.15& 91.69$\pm$0.23&92.95$\pm$0.05&94.87$\pm$0.07\\
 & $P$ &- &-&-& 54.04$\pm$0.34&-&-\\
\bottomrule
\end{tabular}}
\label{link}

\end{table*}

\subsubsection{Impacts of the Prompt Tuning on Privacy Risks}
One of the most significant contributions of GPL compared to traditional GNNs is the prompt mechanism. By comparing \textbf{w/o Prompt} with the other five GPL methods, we explore whether the presence of the prompt mechanism increases the risk of privacy disclosure. However, we found that the high attack success rate in \textbf{w/o Prompt} was similar to, or even slightly higher than, the Prompt method. This finding indicates that the prompt mechanism itself does not inherently increase the privacy risk in graph learning; rather, privacy vulnerabilities stem from the rich structural and attribute information encoded within GNNs. The purpose of the prompt mechanism is to achieve task-specific adjustments using a fixed pre-trained model, not to alter the core representation of node embeddings.

In summary, while prompts can effectively guide downstream tasks, they do not exacerbate privacy risks compared to traditional GNN methods. Nonetheless, it is important to recognize that when sharing trained prompts, they are themselves still vulnerable to attribute inference attacks. There is no inherent conflict between the two.

\subsection{Variable Factors Analysis}
\mypara{Influence of Different Pretraining Models} The `Pretraining, Prompts' paradigm, which has been widely applied in GPL, raises an important question: do different pre-training models lead to varying levels of privacy risks in GPL? Using the \textbf{All-in-One} method as a representative example, Figure \ref{fig:a-p} illustrates the performance of attribute inference attacks across different pre-training models on both the homophilous Cora dataset and the heterophilous Actor dataset (a full comparison of all prompt models' attack results is provided in the \textbf{Appendix}). Notably, we observed that the impact of pre-training models on attack performance does not follow a consistent pattern across datasets. The effectiveness of pre-training methods varies significantly between Cora and Actor. For example, while GraphCL posteriors achieve relatively high AUC on the Actor dataset, they perform poorly on the Cora dataset.

To explore the reasons for this, we also present the performance of the graph prompt models themselves(Figure \ref{fig:ori}), which which accords with the data stated in ProG\cite{zi2024prog}. Similar to the attack results, no clear trend emerges regarding model performance across different pre-training methods.  We hypothesize that this inconsistency stems from the distinct objectives of the pre-training techniques, which are optimized for varying tasks (e.g., contrastive learning, node reconstruction).  As a result, their performance in mitigating privacy attacks differs substantially across datasets and tasks, making it difficult to generalize their effectiveness.  In conclusion, none of the pre-training methods demonstrated consistent resistance to privacy attacks across all datasets and prompt models.  This indicates that, in practical applications, pre-training and prompting methods must be carefully selected or adapted based on the specific characteristics of the dataset and task in order to achieve an optimal balance between model performance and privacy protection.

\begin{figure}
    \centering
\includegraphics[width=0.95\linewidth]{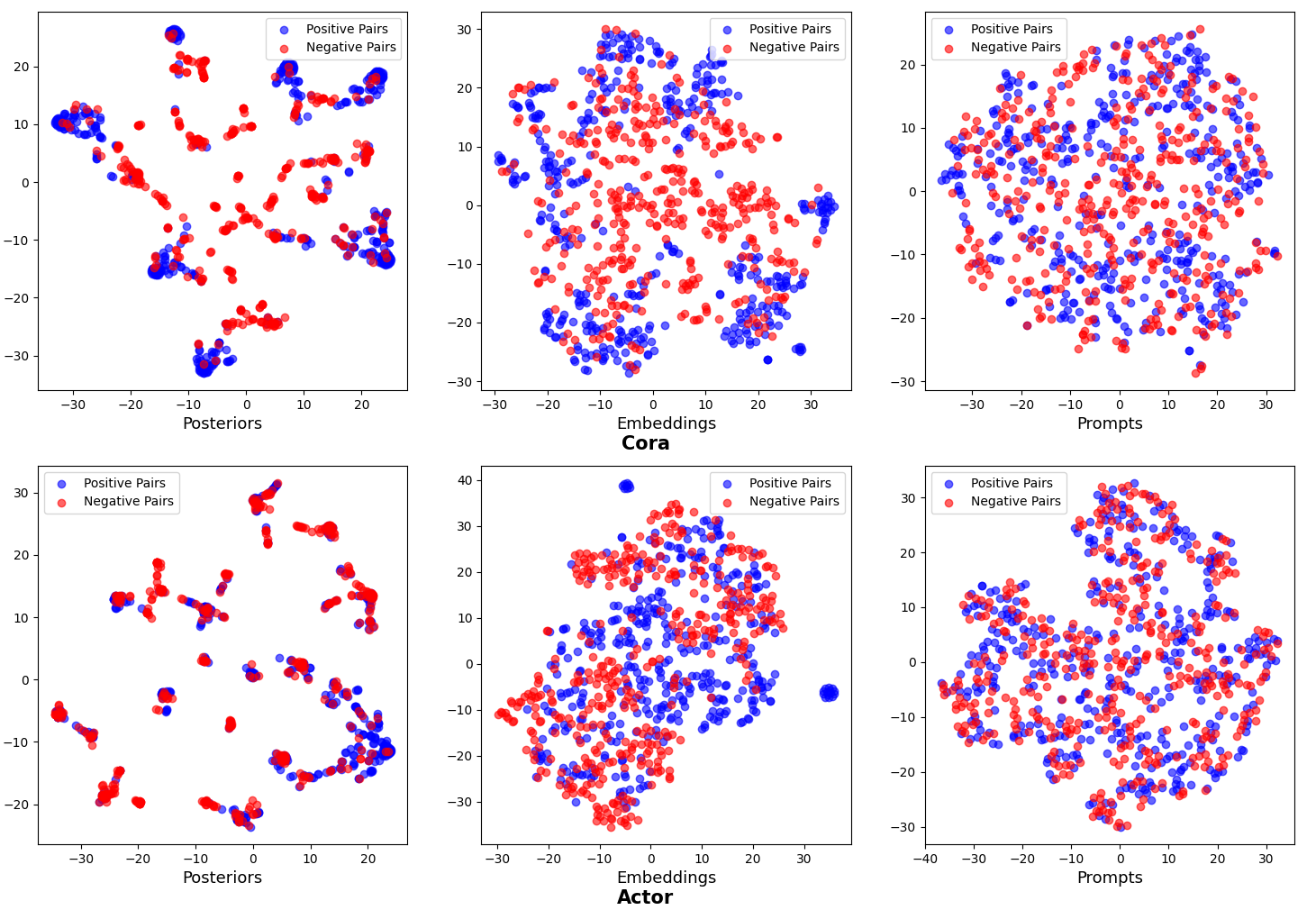}
    \caption{The output of the attack model for 400 randomly sampled positive node pairs and 200 randomly sampled negative node pairs is projected into two-dimensional space using t-SNE.}
    \label{fig:tsne}
\end{figure}

\begin{figure}
    \centering
\includegraphics[width=0.95\linewidth]{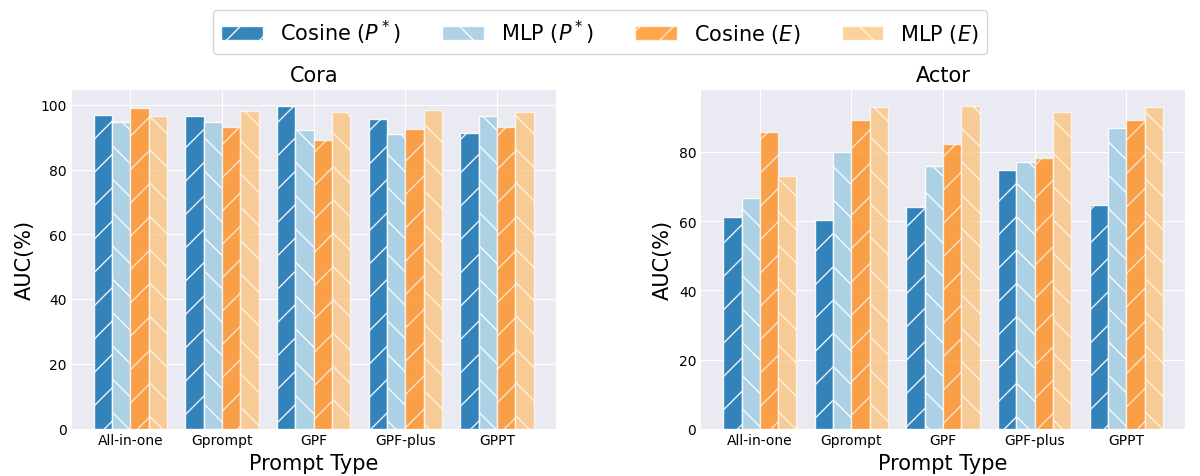}
    \caption{$\mathcal{A}_1$ and $\mathcal{A}_2$ performance under three embeddings}
    \label{fig:Acom}
\end{figure}

\mypara{Influence of k-shots settings} GPL features a flexible prompt design that demonstrates significant potential in sparse data scenarios, commonly known as few-shot settings. To explore different $k$-shot settings, we evaluated whether the value of $k$ (i.e., the number of labeled examples per class) significantly influences the effectiveness of inference attacks. In our experiments, we analyze attribute inference attacks across varying $k=\{1, 5, 10, 50, Full\}$ (see Figure \ref{fig:att-k}) and evaluate the model's intrinsic performance (see Figure \ref{fig:ori-k}). Interestingly, contrary to our initial hypothesis, increasing the $k$ value (the number of samples per class) did not significantly enhance the effectiveness of attacks on node embeddings and posterior outputs. Similarly, although the model's overall performance showed a slight upward trend, it exhibited unstable fluctuations across different $k$ values, aligning with findings in PorG research. 

Our analysis suggests that this phenomenon may result from the model’s ability to adequately represent features through the global capture of graph structure and prompt information, even when $k$ is small. This information saturation effect limits the incremental impact of increasing sample size on privacy disclosure risks. Additionally, prompt learning enhances model robustness, making it less sensitive to perturbations in individual nodes and therefore more resistant to attacks. Furthermore, the varying performance of different models as $k$ changes indicates that certain prompt learning methods achieve strong performance at low $k$ values, with attack effectiveness stabilizing as $k$ increases. These combined factors contribute to the observed stability in attack success rates and privacy risks as $k$ grows.

\section{Link Inference Attacks}
\subsection{Measurement Methodology}
The core objective of a Link Inference Attack (LIA) is to determine whether a link exists between two nodes, $(u,v)$, using information about node pairs. Since Graph Neural Networks (GNNs) aggregate information from neighboring nodes, two connected nodes should be closer to each other in any learned representation space. Consequently, the simplest attack method, similarity-based attack model, involves an unsupervised approach where the distance between node pairs is calculated to predict the existence of a connection. Additionally, we propose a supervised learning method, leveraging a shadow dataset controlled by the attacker, to enhance attack accuracy.

\begin{itemize} 
\item \textbf{Cosine Similarity-Based Attack Model}: This unsupervised link inference method leverages cosine similarity, an efficient metric for measuring proximity between nodes in embedding space. The attacker calculates cosine distance between all node pairs, defined as $d(f(u),f(v)) = d(f(v), f(u))$. By applying a manual threshold, links are inferred when similarity exceeds this value. This straightforward method does not require labeled data, making it advantageous for attackers in scenarios with limited labeled examples.

\item \textbf{MLP-Based Attack Model}:This approach uses a basic binary classification model, with a 3-layer Multilayer Perceptron (MLP) as the baseline for the link inference attack. The MLP comprises hidden layers with $64$ and $32$ neurons, using ReLU activation. The model is trained and tested on a shadow dataset controlled by the attacker, integrating both positive and negative node pair examples. This supervised learning approach improves attack accuracy by leveraging patterns in the shadow dataset, enhancing the model’s ability to distinguish between linked and unlinked node pairs.

\end{itemize}

\subsection{Measurement Settings}
We maintain the same experimental setup as the attribute inference attack in the previous section, but it should be noted that in order to ensure the uniformity of positive and negative samples in the link inference attack. In each experiment, we sample the same number of negative samples as positive samples (link relation) for training and evaluation. Specifically, we first select all linked node pairs, then randomly sample an equivalent number of unlinked node pairs. This negative sampling method follows standard practices in the link prediction literature\cite{he2020stealinglinksgraphneural}. 

\subsection{Measurement Results}
\subsubsection{Evaluations of Attack Models}
We first evaluated the performance of the two proposed attack methods, cosine similarity and MLP, across various datasets. As shown in Figure \ref{fig:Acom}, the attacker leverages the obtained node information to infer links via cosine similarity and machine learning techniques on both the homophilous dataset Cora and the heterophilous dataset Actor. In homophilous graph data, as demonstrated by the inter-class relation matrix in Figure \ref{fig:subfigures}, nodes within the same category exhibit stronger correlations. Consequently, the posterior or embedded representations learned through node classification tasks with graph prompts tend to share similar properties. Thus, on the Cora dataset, tow attack models demonstrate comparable attack performance. In contrast, heterogeneous graph data is characterized by multiple types of information sources, where different nodes and edges carry varied information, resulting in increased structural complexity. This complexity leads to significantly reduced performance for cosine similarity-based inference attacks. In such scenarios, training a shadow dataset with machine learning techniques yields more effective attack outcomes.  

Based on the above findings, in order to ensure the fairness of subsequent comparisons, we mainly report the effects of MLP based attacks.
\subsubsection{Evaluations of different attacker capabilities}
Table \ref{link} presents the LIA results under three attacker capabilities across six datasets.

\mypara{Vulnerability of Node Posteriors to LIAs} When an attacker gains access to posterior information, the success rates of link inference across different graph prompt methods show only minimal variation but are strongly influenced by the dataset characteristics. Specifically, in homophilous datasets, the \textbf{GPF-plus} and \textbf{GPPT} methods consistently achieve higher success rates than other approaches. Additionally, link inference success rates are generally higher in homophilous graph datasets, whereas in heterophilous datasets, they exhibit much greater fluctuation, reflecting the inherent structural diversity and complexity of these graphs.

\mypara{Privacy Risks of Node Embeddings and Prompts} Similar to attribute inference attacks, the success rate of link inference is considerably higher when attackers have access to node embeddings. In many datasets, embedding-based attacks achieve success rates close to or exceeding 90\%. For example, on the Cora dataset, the success rate of embedding-based attacks reaches as high as $98.22\%$ (\textbf{GPF-plus}) and $97.90\%$ (\textbf{Gprompt}), indicating that embeddings are particularly sensitive under graph prompts, making them highly vulnerable to link inference attacks. 

However, we find that when LIA is applied to the prompt of \textbf{GPF-plus}, the degree of privacy leakage was significantly lower than that of attribute inference attacks, especially in heterogeneous data sets, they are basically lower than 70\%. This outcome is tied to the mechanism by which \textbf{GPF-plus} generates prompts, focusing on feature enhancement, which relies more heavily on node features and, as a result, carries less structured information. To further display this, we visualized the distribution of positive and negative sample pairs for the three types of information in \textbf{GPF-plus} (Figure \ref{fig:tsne}). In the Posteriors and Embeddings visualizations, positive and negative node pairs are clearly distinguishable, with the points distributed in separate regions. However, for Prompts, there is significant overlap between positive node pairs (blue dots) and negative node pairs (red dots), indicating that the distinction between the two groups becomes less clear after dimensionality reduction. This highlights the greater challenge of inferring link relationships in heterogeneous datasets using prompt-based methods.

\subsubsection{Impacts of the Prompt Tuning on Privacy Risks}
The attack on \textbf{w/o Prompt} basically reaches the same level as the learning of the five prompts, especially compared with \textbf{GPF}, even presents a more stable attack efficiency. We believe that this is due to the nature of GNN, where the GNN model mainly relies on the connection structure between nodes and the propagation of node features, which may retain a lot of local information at the time of embedding generation. For example, in information aggregation, each node expression can reflect the characteristics of the surrounding neighborhood by capturing the information of the node neighborhood. The GPL also preserves portions of GNN, so in a link inference attack, an attacker can use these aggregation features to reconstruct or infer connections between nodes, rather than relying solely on the information generated by the prompt. 

This finding is consistent with what we have observed in the AIA, that the prompt mechanism does not increase the risk of privacy breaches expressed by its own nodes.

\begin{figure}
    \centering
\includegraphics[width=0.95\linewidth]{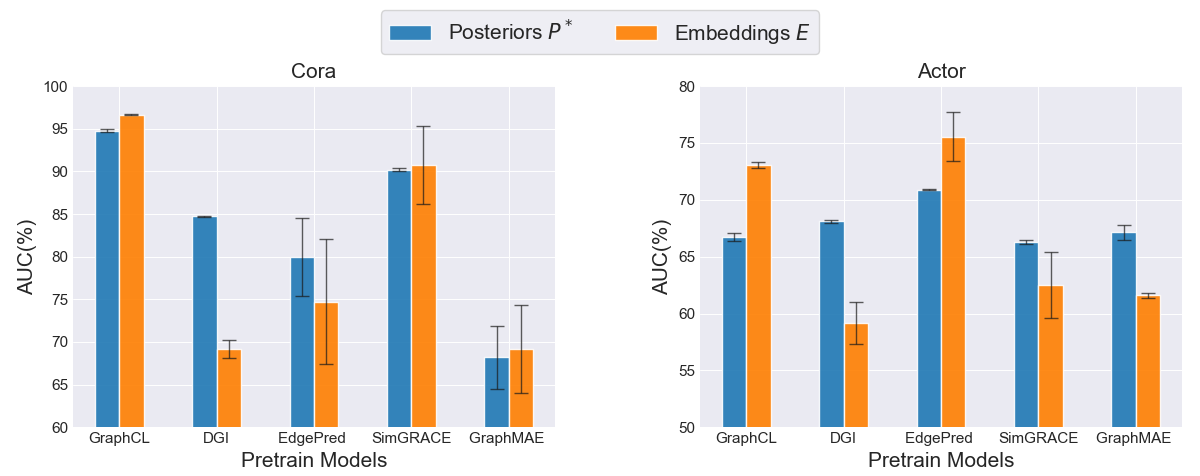}
    \caption{Link inference of attack effects under different pretraining methods (take 'All-in-one' as an example)}
    \label{fig:edge-p}
\end{figure}

\begin{figure}
    \centering
    \includegraphics[width=0.95\linewidth]{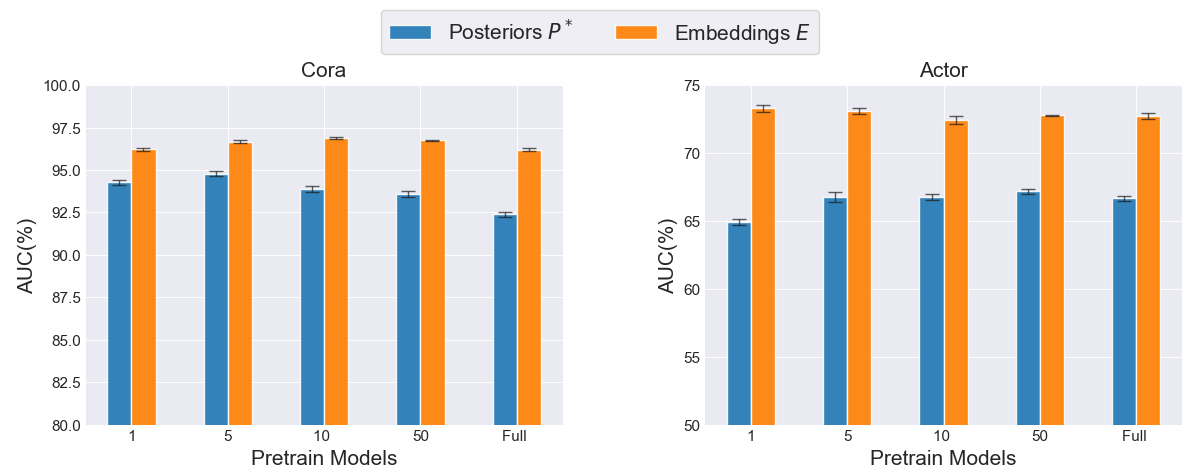}
    \caption{Link inference of attack effects under different k-shot settings (take 'All-in-one' as an example)}
    \label{fig:edg-k}
\end{figure}

\begin{figure*}
    \centering
    \includegraphics[width=0.95\linewidth]{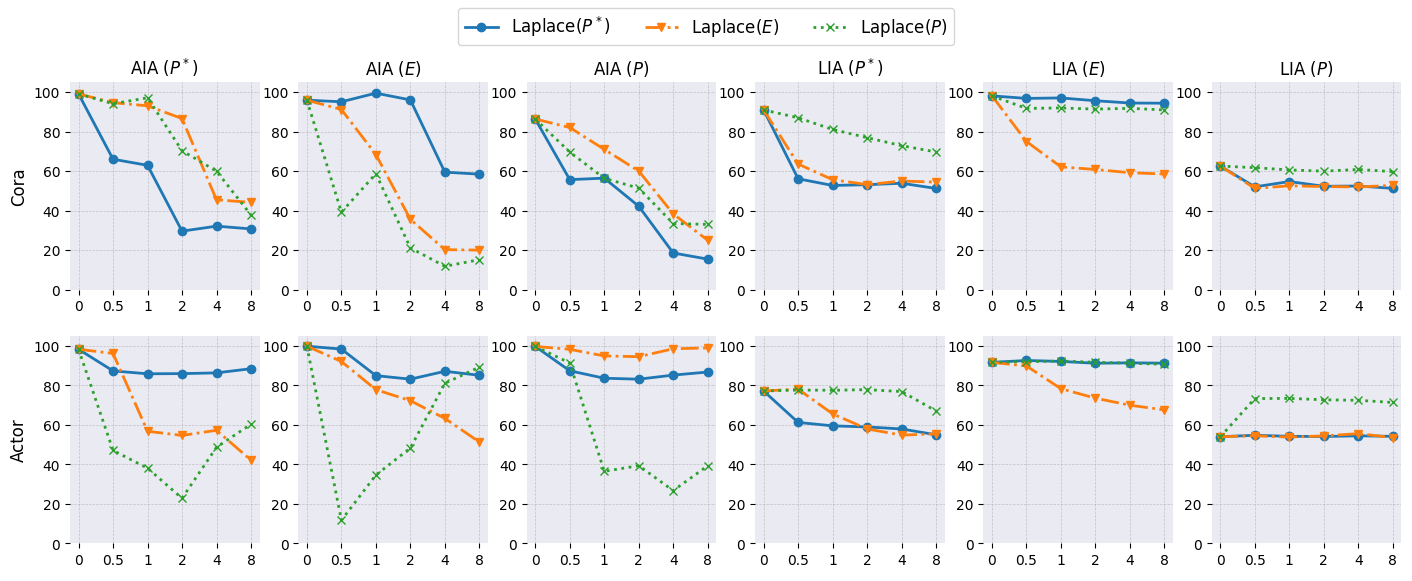}
    \caption{The performance of Attribute inference Attack and Link inference Attack when different amounts of noise are added to the three embeddings respectively}
    \label{fig:noise}
\end{figure*}

\subsection{Variable Factors Analysis}
\mypara{Influence of Different Pretraining Models and k-shots settings} Maintaining the same parameters as in the attribute inference attacks, Figures \ref{fig:edge-p}
 and \ref{fig:edg-k} present the results of the link inference attacks (LIA). Similar to the attribute inference attacks, the performance of LIA shows variability across different pretraining models with no consistent pattern. For example, in the Cora dataset, GraphCL and SimGRACE show a higher attack AUC for embeddings than for posteriors, while in the Actor dataset, DGI achieves a similar effect. This suggests that certain pretraining models may inherently expose more information through embeddings, depending on the dataset structure. However, across various $k$-shot settings, the LIA results remain relatively stable, especially in homogeneous graph data such as Cora. This stability indicates that increasing the sample size $k$ does not significantly impact privacy breach risks in link inference attacks, suggesting that even with limited labeled data, the attacker can maintain a high success rate.

Overall, the attack success rate of LIA is higher than that of Attribute Inference Attacks (AIA) across datasets. This is likely due to the nature of the Graph Prompt Learning (GPL) framework, where prompts and embeddings are designed to capture relational information between nodes to enhance task performance. In such a setting, link relationships are preserved more explicitly, allowing attackers to exploit this structural information effectively. Thus, LIA presents a more potent threat in GPL-based models, as the framework’s design inherently favors the retention of link information in the learned representations.

\section{Defense}
\begin{figure}
    \centering
    \includegraphics[width=0.95\linewidth]{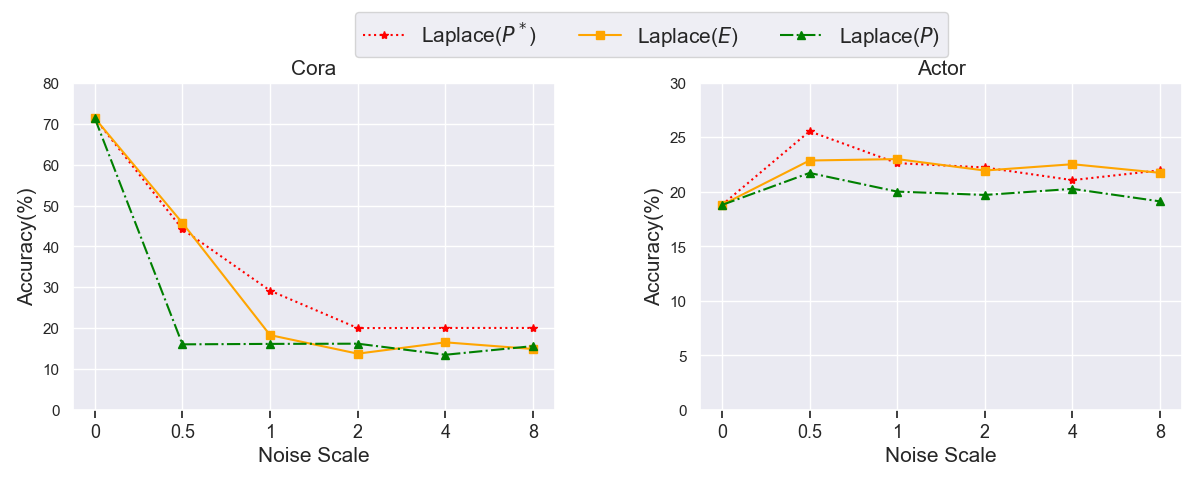}
    \caption{The effect of adding noise on model performance}
    \label{fig:ori-change}
\end{figure}

\begin{figure*}
    \centering
    \includegraphics[width=0.95\linewidth]{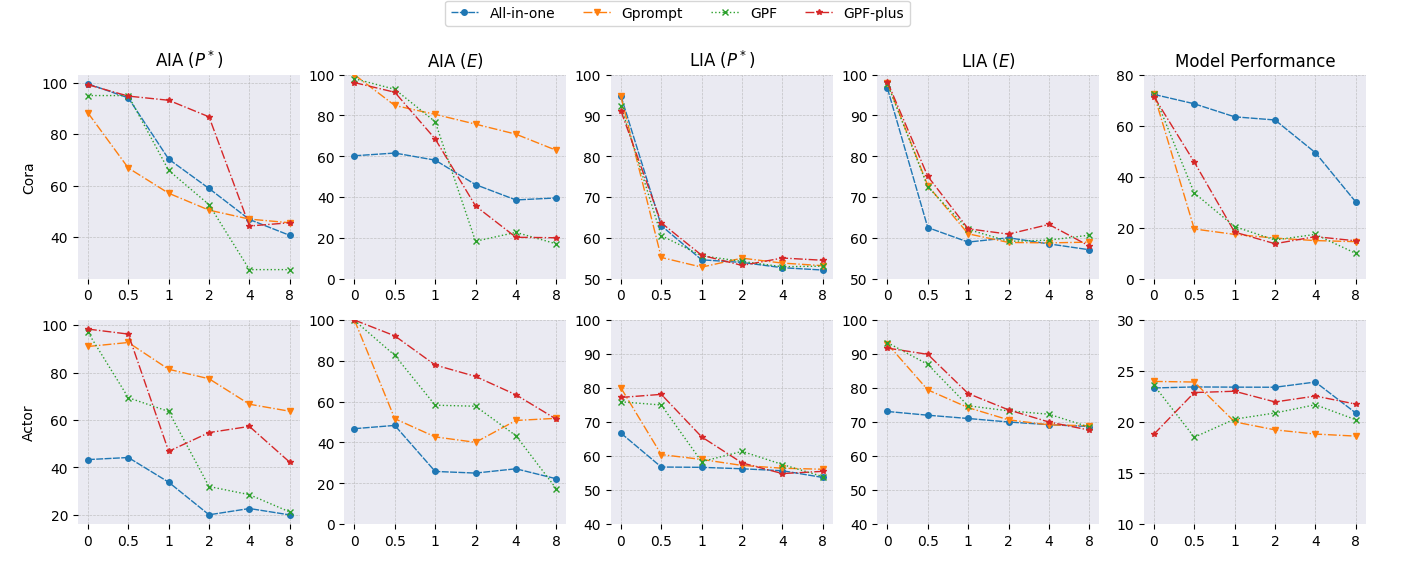}
    \caption{The performance of Attribute inference Attack and Link inference Attack when different amounts of noise are added to the node embeddings}
    \label{fig:emb}
\end{figure*}

\mypara{Laplace Perturbation of Nodes} A common defense mechanism against inference attacks is the addition of noise to the model’s output. Among the various techniques available, Laplacian noise is frequently employed. However, where to add noise is a question worth exploring. In this paper, we attack the node posteriori, node embedding, and prompt respectively according to the ability of the attacker. Therefore, we recommend adding perturbations to each of the three model outputs to compare the impact on the model's defense and raw performance at this time. In particular, for each group of embedded $X$ node, where $X\in \{P^*, E, P\}$, the perturbation in $X_{\text{pri}}$ is defined as:
\begin{equation}
    X_{\text{pri}}=X+\text{Lap}(\beta),
\end{equation}
where $\text{Lap}(\beta)$ is a random variable sampled from the Laplace distribution with scaling parameter $\beta$. Mathematically, this is expressed as $Pr[\text{Lap}(\beta) = x] = \frac{1}{2\beta} e^{-\frac{|x|}{\beta}}$. This method enables us to assess the effectiveness of Laplace perturbation as a defense strategy against various types of inference attacks while also examining its effect on the model's performance.






\mypara{Defense Evaluation Results} In Figure \ref{fig:noise}, we introduce Laplacian noise to the node posterior, embedding, and prompt, respectively, at various noise levels $\beta \in \{0, 0.5, 1, 2, 4, 8\}$, assessing the impact on attribute inference attack (AIA) and link inference attack (LIA) accuracy. Here, $\beta=0$ represents the baseline with no added noise. Experimental results reveal a clear trade-off between privacy protection and model performance across these different noise addition strategies.

First, adding noise to prompts introduces considerable randomness and instability, particularly in heterogeneous graph datasets. Our analysis indicates that GPF-plus enhances node features by training independent embeddings that are specifically tailored to downstream tasks, rather than relying on the graph's inherent structure. As a result, in heterogeneous graphs, aligning these features becomes increasingly challenging, leading to heightened model uncertainty and reduced stability in privacy protection when noise is applied to prompts.

Second, adding noise to both node posteriors and embeddings significantly reduces attack accuracy, particularly at higher noise levels (e.g., $\beta \geq 2$), thereby offering strong privacy protection. However, for link inference attacks and in heterogeneous graph datasets, adding noise to posteriors alone does not provide adequate privacy protection for embeddings. In contrast, embedding-based noise addition proves more robust and adaptable across scenarios, though it may introduce slight privacy risks, especially with sensitivity to prompt-based information. Figure \ref{fig:emb} also illustrates the defense capability and performance of each model after adding noise to node embeddings across several other graph prompt methods. As noise levels increase, the accuracy of both attribute inference and link inference attacks gradually declines, with a notable drop as $\beta$ increases from $0.5$ to $1.0$.

Additionally, as noise levels rise, we observe a significant reduction in model accuracy on downstream tasks (\ref{fig:ori-change}), especially on the Cora dataset, while the Actor dataset is comparatively less impacted. This may be attributed to homogeneous dataset-specific characteristics: Cora's graph structure is closely aligned with node attributes, making it more sensitive to noise perturbations, whereas the Actor dataset, with a more independent structure, maintains performance under high noise conditions. Moving forward, we aim to further investigate privacy protection strategies in graph prompt learning, striving to achieve an optimal balance between privacy and model effectiveness.

\mypara{Takeway} In this study, we investigate the effectiveness of adding Laplace noise to enhance privacy protection in graph prompt learning.  The findings reveal a clear trade-off between privacy and model performance.  Notably, adding noise to prompts increases model instability, particularly in heterogeneous graph datasets, underscoring the need for defense mechanisms tailored to specific dataset characteristics.  While adding noise to node posteriors and embeddings significantly reduces attack accuracy, relying solely on posteriors for privacy protection is insufficient.  Noise added to embeddings demonstrates greater adaptability across scenarios.  As noise levels rise, we observe a marked decline in model accuracy on downstream tasks, especially in the Cora dataset, highlighting the importance of balancing privacy protection with model performance.  Moving forward, we aim to investigate privacy measures in graph prompt learning further to achieve an optimal balance between privacy and efficacy, offering insights for practical applications.

\section{Related Work}
\label{section:related_work}
In this section, we focus on existing research related to Graph Prompt Learning and inference attacks targeting Graph Neural Networks (GNNs).

\mypara{Graph Prompt Learning}
As model complexity and the number of parameters increase, prompt engineering\cite{ye2024promptengineeringpromptengineer, lester2021powerscaleparameterefficientprompt} has emerged as a critical technique in machine learning. Its main goal is to improve model performance and interpretability by designing well-structured prompts. Unlike fine-tuning\cite{chhabra2019data, howard2018universallanguagemodelfinetuning, guo2018spottunetransferlearningadaptive}, which modifies model parameters through backpropagation for specific downstream tasks, prompt engineering focuses on optimizing prompt-related parameters, thereby reducing computational and storage costs. Prompt-based tuning methods are widely used, especially in adapting pre-trained language models to downstream tasks in natural language processing\cite{sahoo2024systematicsurveypromptengineering, li2021prefixtuningoptimizingcontinuousprompts, liu2021pretrainpromptpredictsystematic} and computer vision\cite{bahng2022exploringvisualpromptsadapting, bar2022visualpromptingimageinpainting, huang2023diversityawaremetavisualprompting}. Recently, researchers have extended the concept of prompt learning to graph data\cite{10.1145/3534678.3539249, liu2023graphprompt, 10.1145/3580305.3599256, NEURIPS2023_a4a1ee07, zhu2023sglptstronggraphlearner, 10.1145/3589334.3645685, chen2023ultradpunifyinggraphpretraining, Gong2023PromptTF}. For example, GPPT\cite{10.1145/3534678.3539249} was among the first to explore graph prompt learning, connecting pre-training and downstream node tasks via a link prediction task. Similarly, GraphPrompt\cite{liu2023graphprompt} unifies upstream and downstream tasks through a subgraph task framework, introducing a prompt token to extend graph prompt learning to multiple tasks. The All-in-One method\cite{10.1145/3580305.3599256} further extended the prompt to the graph level, unifying task abstraction at different graph levels into graph-level predictions. GPF\cite{NEURIPS2023_a4a1ee07} proposed a pre-training strategy independent of specific pre-trained models, making it adaptable to various graph structures. Further advancements have been made in heterogeneous graphs\cite{Yu2023HGPROMPTBH}, enhancing graph structures\cite{ge2023enhancing}, and improving pre-trained models\cite{Yu2023MultiGPromptFM}. These developments indicate a growing interest in harnessing prompt learning for complex graph data scenarios.

\mypara{Attribute Inference Attacks}
The goal of attribute inference attacks is to deduce sensitive information about users from publicly available data, such as recommendation systems or other observable outputs. In the context of GNNs, attackers often aim to infer sensitive node attributes from the model’s outputs or node embeddings. Recently, several efforts have been made to extend attribute inference attacks to the graph level. For instance, \cite{zhang2022inference} demonstrates that even in a black-box setting, an attacker can infer sensitive attributes through graph embeddings by leveraging adversarial knowledge and commonly shared attributes. Building upon this, \cite{wang2022group} further extended this idea, showing that incorporating both node attributes and link information into node embeddings could amplify the potential for sensitive data exposure. These findings emphasize the necessity for robust privacy measures in GNN applications.

\mypara{Link Inference Attacks}
Link inference attacks have become a critical challenge in maintaining online privacy. These attacks aim to infer the existence of sensitive links in a network, such as undisclosed social connections or private financial transactions, posing significant risks to user privacy. Attackers typically exploit node similarities within the network, which may arise from node attributes or network structural features such as common neighbors\cite{Grover2016node2vecSF, 10.5555/3327345.3327423}. Traditional link inference methods focused on either node attribute similarity or network structural similarity. However, with the advancement of Graph Representation Learning, recent works increasingly utilize node embeddings to infer links. These embeddings incorporate node attributes, network structure, and auxiliary information, providing a more comprehensive basis for attackers to infer private links\cite{he2020stealinglinksgraphneural,Backes2017walk2friendsIS}. As GNNs become more prevalent, privacy attacks targeting GNNs have gained attention.  For example, Wu et al.\cite{9833806} explored a scenario where, given the server’s access to the graph’s topology while the client provides only node features and labels, an attacker could infer private links in the graph by crafting malicious queries. Zhang et al.\cite{277160} proposed a graph reconstruction attack, aiming to reconstruct a graph similar to the training graph using the model’s learned embeddings. He et al.\cite{he2020stealinglinksgraphneural} systematically analyzed link inference attacks, categorizing them across three dimensions—attacker's background knowledge, attack strategies, and outcomes—culminating in a comprehensive classification of eight distinct types of link-stealing attacks.


\section{Limitation and Future Work}

While our study offers the first comprehensive assessment of privacy risks in GPL, several limitations merit further exploration.

\mypara{Limited Attack Scenarios} This work primarily addresses two inference attacks—Attribute Inference Attack (AIA) and Link Inference Attack (LIA).    Although these reveal significant vulnerabilities, other risks, like subgraph-specific attacks or broader graph-level threats, remain unexamined.    Additionally, complex adversarial scenarios, such as membership inference\cite{He2021NodeLevelMI}, reconstruction attacks\cite{10.1145/3448891.3448939},  and extraction attacks\cite{10.1145/3488932.3497753} are unexplored.    Future research should expand attack models to cover a wider range of privacy threats, especially those in advanced adversarial settings.

\mypara{Defense Mechanism Trade-offs} Laplacian noise perturbation shows promising results in reducing the success of inference attacks. However, as demonstrated in our experiments, increasing noise levels can degrade the performance of downstream tasks, particularly in simpler or homophilous datasets.   Finding an optimal balance between privacy protection and task performance remains a challenge.   Future work could explore more sophisticated defense mechanisms that preserve task performance while providing stronger privacy guarantees.  More privacy-preserving mechanisms\cite{DBLP:journals/corr/abs-2108-04417} may offer potential avenues for enhancing the robustness of graph prompt models against inference attacks.

\mypara{Scalability and Generalization} Our experimental results are based on six real-world datasets and several widely used GPL models.   While these datasets and models cover a range of graph structures, from homophilous to heterophilous graphs, the scalability and generalization of our findings to larger, more complex datasets and models require further validation. In the future, we will expand to conduct privacy security assessments in more real-world data sets.


\section{Conclusion}
This paper is the first to examinesthe privacy risks associated with the GPL. By simulating AIAs and LIAs, we conduct a systematic privacy evaluation of GPL on node-level for the first time. Our findings show that the information shared by the GPL, whether as a service or shared with three parties, significantly increases the risk of sensitive attributes and relational reasoning, with attack success rates as high as 98\% in some datasets. Although there are significant privacy risks associated with specific types of prompts, such as GPF-plus, our research shows that the prompt mechanism itself does not increase the likelihood of privacy breaches in GNN. To solve these risks, we propose a defense strategy based on Laplacian noise perturbation, which effectively reduces the attack success rate. However, achieving a balance between privacy protection and model performance remains challenging. We believe our findings will highlight the importance of privacy considerations in GPL applications, encouraging further research into robust privacy protection strategies in this emerging field.

\newpage



\bibliographystyle{IEEEtran}
\bibliography{main}
%



\section*{Appendix}
\subsection*{A  Notations}
The key notations frequently used throughout this paper are summarized in Table \ref{table:notations}.
\begin{table}[H]
\caption{Summary of the notations used in this paper.}
\centering
\vspace{-0.1cm}
\resizebox{0.9\linewidth}{!} 
{
\begin{tabular}{l|l}
\toprule 
\textbf{Notation} & \textbf{Description} \\ \toprule
	  $\mathcal{G} = \{V,A,X\}$   & Graph   \\ 
      $u, v \in \mathcal{V}$ & Nodes in $\mathcal{G}$ \\ 
      $X \in \mathbb{R}^{n \times d_{X}}$ &  Attributes associated with $u$ \\ 
      $\mathcal{N}_{u}$& Neighborhood nodes of $u$ \\ 
      $P$ & Learnable representation of the prompt \\
	 $P^*$ & Node posterior output\\
      $E$& Node embedding representation\\ 
      $\mathcal{P}$& Graph prompt module \\
      $\mathcal{M}$ & Pre-trained GNN model \\
      $D_{target},D_{shadow}$ & Target/Shadow datasets\\
      $\mathcal{K}$ & Adversary’s Knowledge and Capability\\
\bottomrule
\end{tabular}
}

\label{table:notations}
\end{table}

\subsection*{B  Additional Experimental Results}
\mypara{Performance of three AIA attacks}In our evaluation of three AIA methods, the performance on node embeddings closely mirrors that of node posteriors. The Random Forest-based method consistently underperforms compared to the other two attack methods, particularly on the Cora dataset. In contrast, the MLP and GraphSAGE-based methods demonstrate superior performance, highlighting their robustness in these attack scenarios.

\begin{figure}[b]
    \centering
\includegraphics[width=0.9\linewidth]{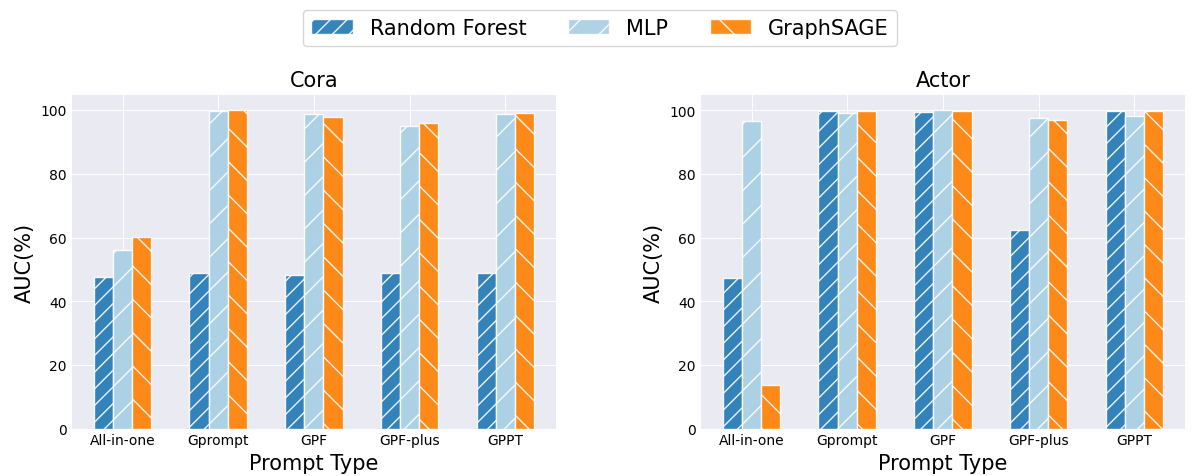}
    \caption{Three models's attack performance under when using posteriors}
    \label{fig:com}
\end{figure}

\mypara{Main results in $10$-shot}We also report the performance of various baseline graph prompt learning methods on AIAs and LIAs using the GraphCL pre-trained model (Tables \ref{a1} and \ref{a2}). Compared with Tables \ref{att} and \ref{link}, changing the shot number $k$ from $5$ to $10$ does not yield significant differences in results. The trends and performance levels remain largely consistent across the two settings, indicating that the number of shots has limited impact on the model’s susceptibility to these attacks.

\begin{figure*}
    \centering
    \includegraphics[width=0.9\linewidth]{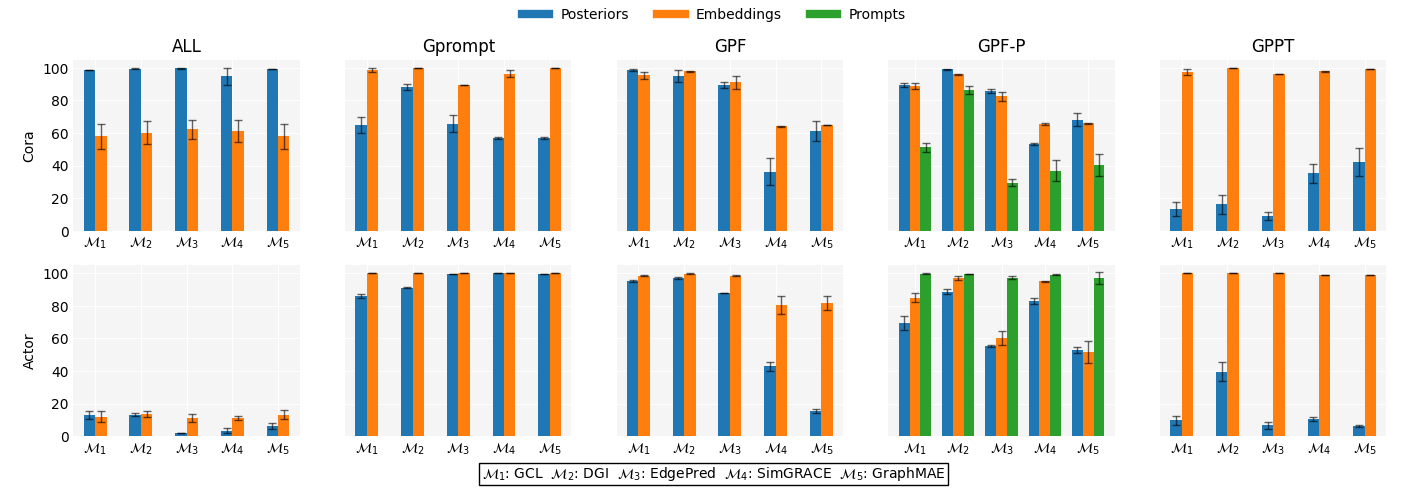}
    \caption{The performance of AIAs in different pretraining GNN models}
    \vspace{-7mm}
    \label{fig:a1}
\end{figure*}

\begin{figure*}
    \centering
    \includegraphics[width=0.9\linewidth]{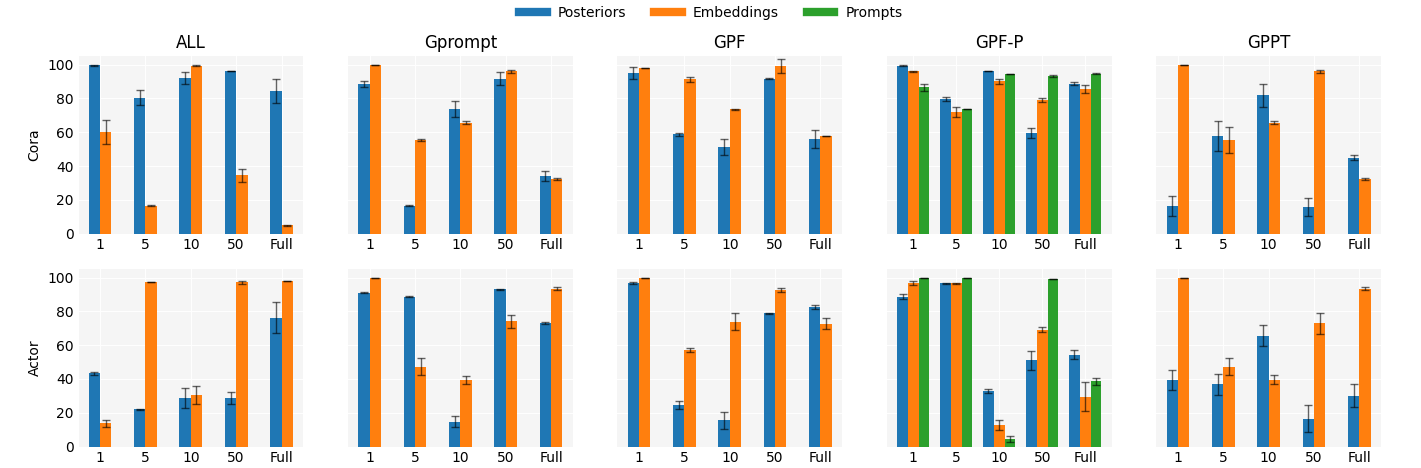}
    \caption{The performance of AIAs in different k-shot settings}
    \vspace{-7mm}
    \label{fig:a2}
\end{figure*}

\begin{figure*}
    \centering
    \includegraphics[width=0.9\linewidth]{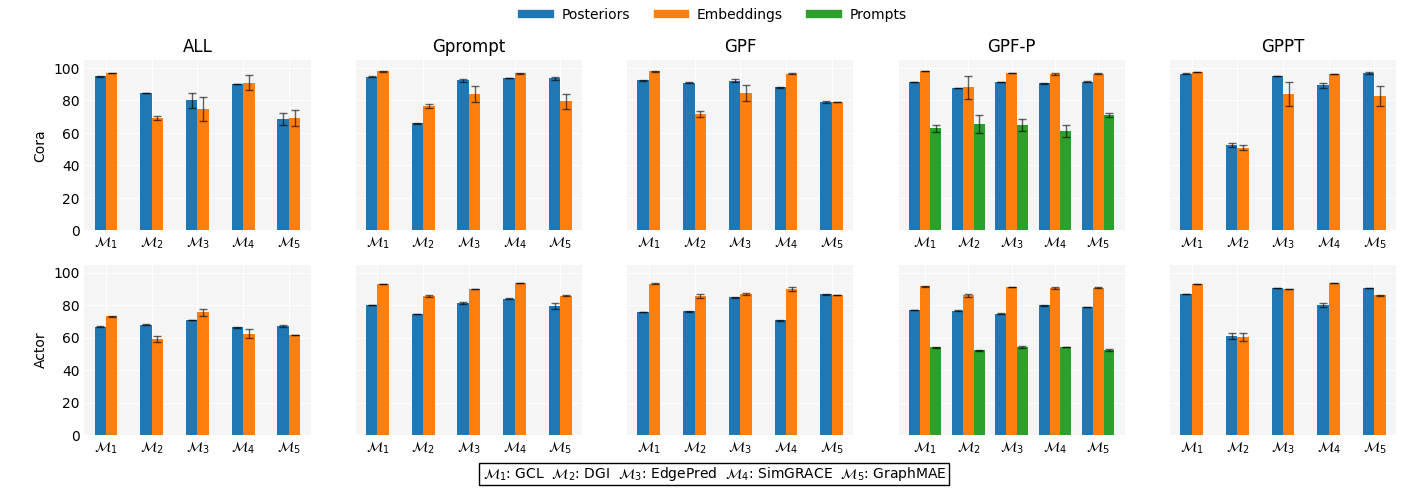}
    \caption{The performance of LIAs in different pretraining GNN models}
    \vspace{-7mm}
    \label{fig:a3}
\end{figure*}

\begin{figure*}
    \centering
    \includegraphics[width=0.9\linewidth]{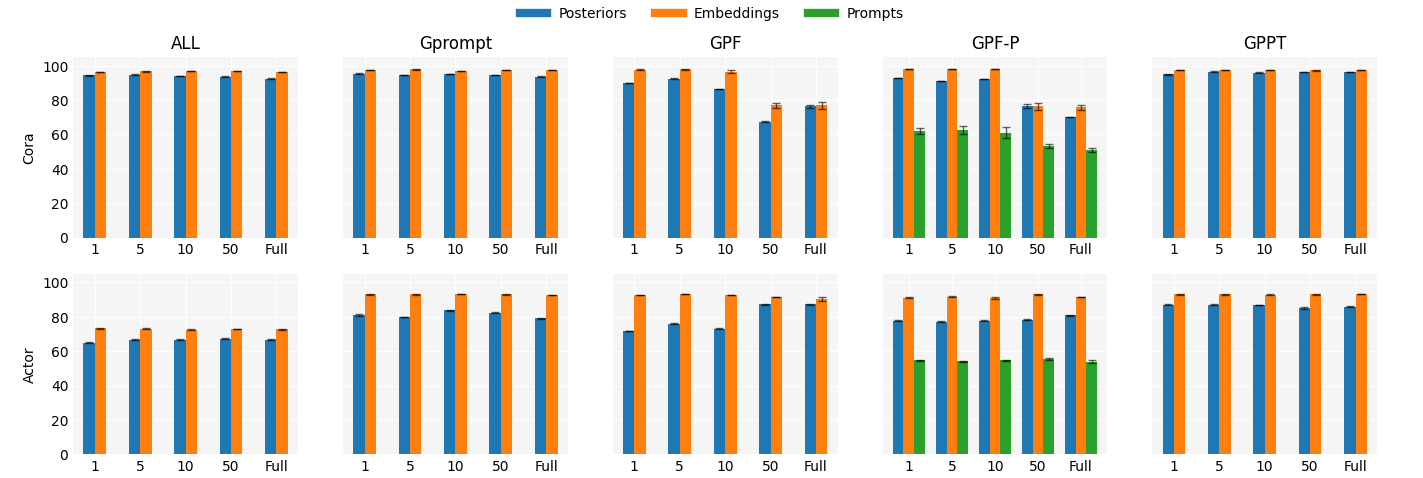}
    \caption{The performance of LIAs in different k-shot settings}
    \vspace{-7mm}
    \label{fig:a4}
\end{figure*}

\mypara{Additional baseline methods}Figure \ref{fig:a1}, \ref{fig:a2},\ref{fig:a3} and \ref{fig:a4} illustrate the attack performance of attribute inference attacks and link inference attacks for different pre-training models and $k$-shot settings in the remaining graph prompt baseline methods. The experimental results show that in most cases, the pre-trained model and the size of $k$ do not have a direct and regular effect on the success of the attack.


\begin{table*}[bt]
\small
\centering
\tiny
\renewcommand\arraystretch{0.8}
\caption{Experimental results of attribute  inference attacks in 10-shot}
\resizebox{0.8\linewidth}{!}{\begin{tabular}{c|c|cccccc}
\toprule
\multirow{3}{*}{\textbf{Dataset}} & \multirow{3}{*}{\textbf{Knowledge
}} & \multicolumn{5}{c}{\textbf{Prompt Type}} \\
\cmidrule{3-8}
&  & All in one & Gprompt & GPF & GPF-plus & GPPT&w/o Prompt  \\
\midrule
\multirow{3}{*}{Cora} & $P^*$ &99.61$\pm$0.03  & 65.65 $\pm$5.14 &89.59$\pm$1.87 &85.94$\pm$1.36 &9.33$\pm$2.51&97.59$\pm$0.02 \\
 & $E$ &62.36$\pm$5.81  &   98.33$\pm$0.02&91.09$\pm$3.80&82.63$\pm$2.71 &96.43$\pm$0.02&94.94$\pm$0.95\\
 & $P$&- & - &- &29.47$\pm$0.02&- &-\\
\midrule
\multirow{3}{*}{Photo} & $P^*$ &78.33$\pm$0.12&84.96$\pm$0.14&82.76$\pm$0.12 & 79.62$\pm$0.28&75.85$\pm$12.31&79.55$\pm$0.26\\
 & $E$ &84.41$\pm$0.23&90.94$\pm$0.13&90.57$\pm$0.17 & 90.46$\pm$0.18&90.94$\pm$0.14&90.73$\pm$0.20\\
 & $P$ &- &-&-& 87.23$\pm$0.33&- &- \\
\midrule
\multirow{3}{*}{Computers} & $P^*$ &76.44$\pm$0.21&84.17$\pm$0.05&80.16$\pm$0.07& 77.42$\pm$0.08&49.99$\pm$0.02&83.49$\pm$0.10\\
 & $E$ &82.18$\pm$0.25 &86.88$\pm$0.28&87.02$\pm$0.22 & 86.51$\pm$0.15&86.93$\pm$0.27&88.56$\pm$0.20\\
 & $P$&- &-&-& 81.69$\pm$0.20&- &-\\
 \midrule
 \multirow{3}{*}{ogbn-arxiv} & $P^*$ &74.89$\pm$0.06 &80.00$\pm$0.05&82.25$\pm$0.24 & 82.09$\pm$0.19&77.52$\pm$0.19&80.80$\pm$0.03\\
  & $E$ &75.63$\pm$0.06 &82.22$\pm$0.48&85.01$\pm$0.29 & 82.01$\pm$0.05&81.19$\pm$0.73&84.29$\pm$0.16\\
  & $P$ &- & - &-& 88.31$\pm$0.17&-&- \\
 \midrule
 \multirow{3}{*}{Squirrel} & $P^*$ &75.04$\pm$0.52 &83.09$\pm$3.27&92.58$\pm$0.17 & 85.63$\pm$0.47&60.36$\pm$7.37&75.15$\pm$0.59\\
 & $E$ &45.76$\pm$1.13 &82.81$\pm$4.82&81.83$\pm$0.79 &86.29$\pm$3.09&79.20$\pm$4.66&77.30$\pm$4.11\\
 & $P$ &- &-&-&97.53$\pm$0.18&-&-\\
 \midrule
 \multirow{3}{*}{Actor} & $P^*$ &2.06$\pm$0.04 &99.47$\pm$0.03&87.83$\pm$0.23 & 55.21$\pm$0.52&6.72$\pm$2.30&99.26$\pm$0.01\\
 & $E$ &11.27$\pm$2.27 &99.93$\pm$0.01&98.54$\pm$0.49 & 60.04$\pm$4.24&99.94$\pm$0.01&99.68$\pm$0.18\\
 & $P$ &- &-&-& 97.26$\pm$0.94&-&-\\
\bottomrule
\end{tabular}}
\label{a1}

\end{table*}

\begin{table*}[ht]
\small
\centering
\tiny
\renewcommand\arraystretch{0.8}
\caption{Experimental results of link  inference attacks in 10-shot}
\resizebox{0.8\linewidth}{!}{\begin{tabular}{c|c|cccccc}
\toprule
\multirow{3}{*}{\textbf{Dataset}} & \multirow{3}{*}{\textbf{Knowledge
}} & \multicolumn{5}{c}{\textbf{Prompt Type}} \\
\cmidrule{3-8}
&  & All in one & Gprompt & GPF & GPF-plus & GPPT&w/o Prompt  \\
\midrule
\multirow{3}{*}{Cora} & $P^*$ &93.87$\pm$0.16  & 94.97 $\pm$0.05 &86.49$\pm$0.14 &92.28$\pm$0.08 &95.96$\pm$0.19&88.60$\pm$0.05 \\
 & $E$ &96.88$\pm$0.08  &   96.77$\pm$0.07&96.45$\pm$0.98&98.13$\pm$0.05 &97.33$\pm$0.10&98.36$\pm$0.06\\
 & $P$ &- & - &- &60.93$\pm$3.09&-&- \\
\midrule
\multirow{3}{*}{Photo} & $P^*$ &88.75$\pm$0.06&93.16$\pm$1.05&90.58$\pm$0.06 & 89.05$\pm$0.11&90.16$\pm$1.39&92.74$\pm$0.05\\
 & $E$ &92.80$\pm$0.70&97.70$\pm$0.05&97.84$\pm$0.10& 97.71$\pm$0.10&97.68$\pm$0.19&98.15$\pm$0.16\\
 & $P$ &- &-&-& 78.60$\pm$0.18&- &- \\
\midrule
\multirow{3}{*}{Computers} & $P^*$ &79.76$\pm$0.12&89.09$\pm$2.11&86.72$\pm$0.06& 86.69$\pm$0.06&67.24$\pm$11.22&89.85$\pm$0.08\\
 & $E$ &82.96$\pm$3.61&95.60$\pm$0.38&96.66$\pm$0.16 & 95.05$\pm$1.15&95.53$\pm$0.36&96.69$\pm$0.04\\
 & $P$ &- &-&-& 75.25$\pm$0.16&-&- \\
 \midrule
 \multirow{3}{*}{ogbn-arxiv} & $P^*$ &89.56$\pm$0.21&94.01$\pm$0.09&93.30$\pm$0.05 & 94.21$\pm$0.13&91.34$\pm$0.22&91.67$\pm$0.19\\
  & $E$ &92.20$\pm$0.07 &95.93$\pm$0.19&98.21$\pm$0.08 & 96.00$\pm$0.57&97.21$\pm$0.05&98.84$\pm$0.15\\
  & $P$ &- & - &-& 75.21$\pm$3.48&-&- \\
 \midrule
 \multirow{3}{*}{Squirrel} & $P^*$ &85.03$\pm$0.07&89.99$\pm$0.05&91.18$\pm$0.05 & 91.10$\pm$0.02&95.31$\pm$0.02&88.99$\pm$0.09\\
 & $E$ &89.42$\pm$0.18&97.58$\pm$0.04&97.23$\pm$0.20 & 96.38$\pm$0.30&97.58$\pm$0.03&97.89$\pm$0.10\\
 & $P$ &- &-&-& 68.67$\pm$0.26&-&-\\
 \midrule
 \multirow{3}{*}{Actor} & $P^*$ &66.74$\pm$0.23 &83.64$\pm$0.14&73.07$\pm$0.18 & 77.77$\pm$0.10&86.78$\pm$0.22&77.47$\pm$0.26\\
 & $E$&72.44$\pm$0.29&93.04$\pm$0.09&92.77$\pm$0.10 & 90.84$\pm$0.50&92.77$\pm$0.16&94.51$\pm$0.06\\
 & $P$ &- &-&-& 54.76$\pm$0.32&-&-
 \\
\bottomrule
\end{tabular}}
\label{a2}

\end{table*}




\end{document}